\documentclass[prl, twocolumn, superscriptaddress, floatfix]{revtex4-2}
\usepackage{graphicx}
\usepackage{float}
\usepackage{amsmath,amssymb,amstext,dsfont,tikz,graphicx,physics,mathtools,bm,
simpler-wick}
\usepackage[colorlinks=true, allcolors=purple]{hyperref}

\usepackage{soul,xcolor}
\setstcolor{red}

\begin{document}

\title{Scrambling Transition in a Radiative Random Unitary Circuit}
\author{Zack Weinstein}
\thanks{Z.W. and S.K. contributed equally to this work.}
\affiliation{Department of Physics, University of California, Berkeley, California 94720, USA}

\author{Shane P. Kelly}
\thanks{Z.W. and S.K. contributed equally to this work.}
\affiliation{Institute for Physics, Johannes Gutenberg University of Mainz, D-55099 Mainz, Germany}

\author{Jamir Marino}
\affiliation{Institute for Physics, Johannes Gutenberg University of Mainz, D-55099 Mainz, Germany}

\author{Ehud Altman}
\affiliation{Department of Physics, University of California, Berkeley, California 94720, USA}
\affiliation{Materials Sciences Division, Lawrence Berkeley National Laboratory, Berkeley, CA 94720, USA}

\date{\today}

\begin{abstract}
We study quantum information scrambling in a random unitary circuit that exchanges qubits with an environment at a rate $p$. As a result, initially localized quantum information not only spreads within the system, but also spills into the environment. Using the out-of-time-order correlator (OTOC) to characterize scrambling, we find a nonequilibrium phase transition in the directed percolation universality class at a critical swap rate $p_c$: for $p < p_c$ the ensemble-averaged OTOC exhibits ballistic growth with a tunable light cone velocity, while for $p > p_c$ the OTOC fails to percolate within the system and vanishes uniformly within a finite timescale, indicating that all local operators are rapidly swapped into the environment. To elucidate its information-theoretic consequences, we demonstrate that the transition in operator spreading coincides with a transition in an observer's ability to decode the system's initial quantum information from the swapped-out, or ``radiated," qubits. We present a simple decoding scheme which recovers the system's initial information with perfect fidelity in the nonpercolating phase, and with continuously decreasing fidelity with decreasing swap rate in the percolating phase. Depending on the initial state of the swapped-in qubits, we further observe a corresponding entanglement transition in the coherent information from the system into the radiated qubits.

\end{abstract}

\maketitle

Understanding the complexity of quantum states and operators undergoing time evolution is a key challenge with potential implications across fields, from condensed matter physics, through quantum gravity, to quantum computation.
In condensed matter physics, insights on the growth of operator complexity have inspired new ways of computing the dynamics of thermalizing systems~\cite{prosen_matrix_2009,white_quantum_2018,xu_locality_2019,zhou_entanglement_2020,rakovszky_dissipation-assisted_2022,von_keyserlingk_operator_2022,kleinkvorning_time_2022,schuster_operator_2022}.
Operator growth, as measured for example by out-of-time-order correlations (OTOCs)~\cite{larkin_quasiclassical_1969,maldacena_bound_2016,swingle_unscrambling_2018,xu_scrambling_2022}, is also considered as key to relating boundary to bulk dynamics in the conjectured AdS/CFT correspondence~\cite{sekino_fast_2008,lashkari_towards_2013,shenker_black_2014,shenker_multiple_2014,roberts_localized_2015,brown_holographic_2016,brown_quantum_2017}. Furthermore, understanding complexity growth in terms of scrambling of quantum information has revealed connections between the dynamics of black holes and the capacity of artificial quantum circuits to encode and process quantum information~\cite{page_average_1993,hayden_black_2007,harlow_quantum_2013,yoshida_efficient_2017}. 

Physical observables evolved by simple models of unstructured unitary circuits or of thermalizing many-body Hamiltonians are expected to scramble and grow in complexity indefinitely, or at least to astronomical timescales~\cite{brown_quantum_2017,parker_universal_2019}. 
But these models may be too simplified in some cases. For example, as information is scrambled in a black hole some of it is lost to Hawking radiation~\cite{hawking_particle_1975}. Similarly decoherence in quantum circuits implies that some of the information is ultimately lost, or shared with external degrees of freedom~\cite{bao_theory_2020,weinstein_measurement-induced_2022,schuster_operator_2022}. Can there exist sharp thresholds or phase transitions in scrambling, or in the flow of quantum information, tuned by the rate of such loss processes?

Recently, it has been discovered that random unitary circuits (RUCs) interspersed with local projective measurements can exhibit two distinct dynamical phases, characterized respectively by the partial protection or rapid destruction of initially encoded quantum information, which are separated by a continuous phase transition at a nonzero critical measurement rate \cite{aharonov_quantum_2000,skinner_measurement-induced_2019,li_measurement-driven_2019,choi_quantum_2020,gullans_dynamical_2020,bao_theory_2020,potter_entanglement_2022,fisher_random_2022}. However, measurements are highly nonrepresentative of generic errors, and moreover, such measurement-induced phase transitions (MIPTs) typically face exponentially large postselection barriers to experimental observation \cite{gullans_scalable_2020,lee_decoding_2022,li_cross_2022}. It is natural to ask if a phase transition in scrambling and information flow can occur in a RUC without measurements, thereby avoiding the postselection problem altogether.

\begin{figure*}[t]
	\includegraphics[width = \textwidth]{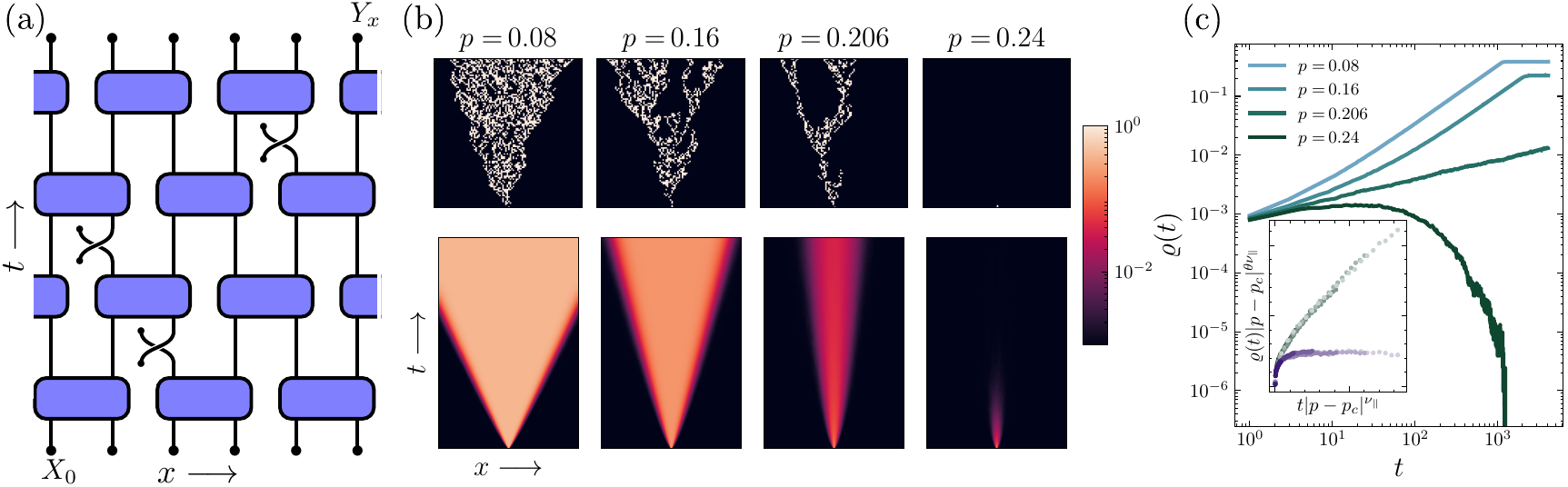}
	\caption{(a) Circuit diagram for the model studied. An initial operator $X_0$ is Heisenberg evolved via a random unitary circuit. In between layers of unitary gates, swaps with ancilla qubits occur with probability $p$. The OTOC $C(x,t)$ is obtained from the commutator $\comm{X_0(t)}{Y_x}$. (b) Top: OTOCs for typical Clifford circuit realizations, for four swap rates $p$, for a system of size $N = 100$. White denotes $C(x,t) = 1$, while black denotes $C(x,t) = 0$. Bottom: averaged OTOC for the same swap rates in a system of size $N=1024$, depicting the narrowing and eventual vanishing of the light cone. Colors are plotted on a log scale for increased contrast. (c) Integrated OTOC $\varrho(t) = \frac{1}{N} \sum_x \overline{C(x,t)}$ for several swap rates $p$ in a system of size $N=1024$. $\varrho(t)$ exhibits linear growth and saturates at a finite value for $p < p_c$, exhibits power-law growth with exponent $\theta \simeq 0.3175$ at the critical point $p_c \simeq 0.206$, and rapidly decays to zero for $p > p_c$. Inset: scaling collapse for several swap rates below $p_c$ (green) and above $p_c$ (purple), using DP exponents $\theta_{\text{DP}} \approx 0.3136$ and $\nu_{\parallel,\text{DP}} \approx 1.734$ \cite{hinrichsen_nonequilibrium_2000,jensen_low-density_1999}.}
	\label{fig:fullwidth}
\end{figure*}

In this Letter, we present a simple model of a RUC that exhibits a phase transition from scrambling to nonscrambling dynamics. We extend previous works \cite{nahum_quantum_2017,nahum_operator_2018,von_keyserlingk_operator_2018,rakovszky_diffusive_2018,khemani_operator_2018,chan_solution_2018} exploring operator growth in closed-system RUCs by allowing the system to exchange qubits with an environment \cite{banerjee_solvable_2017,agarwal_toy_2020,piroli_random_2020,maldacena_syk_2021}; as a consequence, initially localized quantum information not only spreads within the system, but also spills into the environment. Using a mapping to a classical nonequilibrium statistical mechanics model of population dynamics, we show that the circuit exhibits tunable scrambling: increasing the rate of qubit swaps reduces the OTOC light cone velocity within the system until it vanishes at a critical swap rate. At this point the model exhibits a phase transition in the directed percolation (DP) universality class \cite{broadbent_percolation_1957,hinrichsen_nonequilibrium_2000,henkel_non-equilibrium_2008,chertkov_characterizing_2022} to a nonscrambling phase. For swap rates above this threshold, all local operators initially within the system are rapidly swapped to the environment.

In contrast to previous works on operator growth in open systems \cite{knap_entanglement_2018,yoshida_disentangling_2019,zhang_information_2019,dominguez_decoherence_2021,dominguez_dynamics_2021,zanardi_information_2021,andreadakis_scrambling_2022,bhattacharya_operator_2022,schuster_operator_2022}, the transition in the OTOC described here requires that an observer has access to the swapped-out, or ``radiated,'' environment qubits. To determine the implications of the transition in operator spreading on the flow of quantum information, we consider a thought experiment in which an observer attempts to recover quantum information stored in the initial state of the system from the radiated qubits alone. Motivated by an analogy to previous studies of quantum information scrambling in black holes \cite{hayden_black_2007,yoshida_efficient_2017,agarwal_toy_2020,piroli_random_2020}, we provide a simple algorithm by which an observer with access to the radiated qubits can decode this quantum information with perfect fidelity in the nonpercolating phase of the circuit, but with an imperfect fidelity set by the DP survival probability in the percolating phase. We numerically demonstrate a corresponding transition in the coherent information into the radiated qubits, which depends nontrivially on the observer's knowledge of the qubits swapped into the system.

\textit{Model.---}We consider a one-dimensional system of $N$ qubits with periodic boundary conditions undergoing brick-wall random unitary evolution [Fig.~\ref{fig:fullwidth}(a)]. Each two-qubit unitary gate is independently drawn from either the Haar or Clifford ensemble. Between layers of unitary gates, each system qubit is swapped with an environmental ancilla qubit with probability $p$; crucially, we use a fresh ancilla qubit for each such interaction, and we do not trace out the ancilla following the system-ancilla interaction. For now, we leave the initial state $\rho_0^{SE}$ on the system $S$ and environment $E$ unspecified.

We study operator spreading in the RUC $U_t$ by computing the out-of-time-order correlator (OTOC)~\cite{larkin_quasiclassical_1969,maldacena_bound_2016,nahum_operator_2018,von_keyserlingk_operator_2018,xu_scrambling_2022}, defined here as
\begin{equation}
\label{eq:otoc}
	C(x,t) = \frac{1}{4} \tr{ \rho_0^{SE} \comm{X_0(t)}{Y_x}^{\dagger} \comm{X_0(t)}{Y_x} },
\end{equation}
where $Y_x$ is the Pauli-$Y$ operator for the $x$th qubit, and $X_0(t) = U_t^{\dagger} X_0 U_t$ is the Heisenberg-evolved Pauli-$X$ operator for the zeroth qubit after $t$ layers of unitary gates.

References~\cite{nahum_operator_2018,von_keyserlingk_operator_2018} previously studied operator spreading in closed-system RUCs using the OTOC. Upon introducing swap gates with an environment, a new physical feature emerges: the operator $X_0(t)$ not only spreads within the system, but also spills into the environment. By tuning the swap rate $p$, the rate of operator growth within the system can be slowed or even halted altogether.

To see this concretely, it is illuminating to consider the evolution of the OTOC in random Clifford circuits \cite{gottesman_stabilizer_1997,gottesman_heisenberg_1998}; see the Supplemental Material \cite{SOM} for a corresponding calculation for Haar-random circuits. Since Eq.~(\ref{eq:otoc}) is second-order in $U_t \otimes U_t^*$ and the Clifford ensemble forms a unitary 3-design, the ensemble-averaged OTOC $\overline{C(x,t)}$ behaves identically in the Haar and Clifford circuits \cite{webb_clifford_2016,zhu_multiqubit_2017}. Whereas $X_0(t)$ evolves into a superposition of many Pauli strings in generic circuits, in a Clifford circuit $X_0(t)$ remains a single Pauli string at all times, and $C(x,t) = 1$ whenever $X_0(t)$ has Pauli content $X$ or $Z$ at site $x$ and vanishes otherwise. Noting that each of the three Pauli operators appear within $X_0(t)$ at a given site with equal probability, we can express $\overline{C(x,t)} = \frac{2}{3} \overline{n_x(t)}$ in terms of an occupation number $n_x(t)$ which equals one whenever $X_0(t)$ has nontrivial (i.e., nonidentity) Pauli content on site $x$ in a given Clifford circuit.

We interpret the evolution of the ensemble $\qty{n_x(t)}$ as a stochastic evolution of particles on a tilted square lattice. Vertices of the lattice sit at the center of unitary gates, and the occupation numbers $\qty{n_x(t)}$ at each integer time step define a distribution of particles along the edges of the lattice. The rules for obtaining $\qty{n_x(t+1)}$ from $\qty{n_x(t)}$ are determined by noting that a two-qubit Clifford gate can evolve a nontrivial two-qubit Pauli string to any of 15 nontrivial two-qubit Pauli strings (up to phase) with equal probability, while the trivial Pauli string always evolves to the trivial Pauli string. Upon adding system-ancilla swap gates, the Pauli content of each site $x$ has an additional probability $p$ of swapping onto an environment qubit: effectively, the particle hops from the system to the environment, and the Pauli content of site $x$ is set to the identity. We therefore obtain the following probabilities for the propagation of particles from an occupied vertex \footnote{A given vertex is considered occupied if at least one of its two incoming edges are occupied.}:
\begin{equation}
\label{eq:dp_rules}
	\tikz{
            \draw[->, line width = 1.0pt] (-1.25, -0.25) -- (-1.25, 0.5);
            \node at (-1.25, -0.5) {$t$};
		\draw[line width = 1.0pt] (0,0) -- (-0.5, 0.5);
		\draw[line width = 1.0pt] (0,0) -- (0.5, 0.5);
		\filldraw (0,0) circle (0.05);
		\node at (0, -0.5) {$\frac{3}{5}(1-p)^2$,};
	} \hspace{0.4em} \tikz{
		\draw[line width = 1.0pt] (0,0) -- (-0.5, 0.5);
		\draw[dotted, line width = 1.0pt] (0,0) -- (0.5, 0.5);
		\filldraw (0,0) circle (0.05);

		\draw[dotted, line width = 1.0pt] (1.1, 0) -- (0.6, 0.5);
		\draw[line width = 1.0pt] (1.1, 0) -- (1.6, 0.5);
		\filldraw(1.1, 0) circle (0.05);
		\node at (0.65, -0.5) {$\frac{1}{5}(1-p) + \frac{3}{5} p (1-p)$,};
	} \hspace{0.4em} \tikz{
		\draw[dotted, line width = 1.0pt] (0,0) -- (-0.5, 0.5);
		\draw[dotted, line width = 1.0pt] (0,0) -- (0.5, 0.5);
		\filldraw (0,0) circle (0.05);
		\node at (0, -0.5) {$\frac{3}{5}p^2 + \frac{2}{5}p$,};
	}
\end{equation}
where a dark line indicates the propagation of a particle, while a dotted line indicates no propagation of a particle. The case $p=0$ returns the results of Refs. \cite{nahum_operator_2018,von_keyserlingk_operator_2018}, which can be interpreted as the stochastic evolution of particles which can diffuse, spread, or coalesce, but never annihilate; the average behavior of the OTOC can be computed exactly in this case, and one finds a ballistic growth of the OTOC with a light cone front that broadens as $\sqrt{t}$. 

In contrast, introducing $p > 0$ allows particles to annihilate, and yields the phenomenology of a DP problem \cite{broadbent_percolation_1957,hinrichsen_nonequilibrium_2000,henkel_non-equilibrium_2008}. The OTOC's light cone velocity continuously decreases with increasing $p$ and vanishes at critical swap rate $p_c$, at which point there is a phase transition in the DP universality class \cite{SOM}. For swap rates $p < p_c$, $X_0(t)$ percolates throughout the system qubits with a finite probability $P(t)$ corresponding to the survival probability of the associated DP process. On the other hand, for $p > p_c$ the particle distribution is rapidly driven to the absorbing state $n_x(t) = 0$; the entire operator $X_0(t)$ is swapped into the environment within a finite timescale, and $C(x,t)$ vanishes uniformly at all times thereafter. These qualitative features are observed in Clifford numerical simulations, as demonstrated in Fig.~\ref{fig:fullwidth}(b). 

Note that the stochastic rules (\ref{eq:dp_rules}) are not microscopically equivalent to standard bond DP, due to correlations between the two edges leaving a vertex. This feature is not expected to affect the universal behavior of the two phases or the transition, in accordance with the DP hypothesis \cite{janssen_nonequilibrium_1981,grassberger_phase_1982}. To confirm that the phase transition lies within the DP universality class, we use Clifford simulations to numerically compute the integrated OTOC $\varrho(t) = \frac{1}{N} \sum_x \overline{C(x,t)}$ for several swap rates [Fig.~\ref{fig:fullwidth}(c)]. We find that $\varrho(t)$ grows linearly and saturates at a nonzero value for $p < p_c$, while it rapidly decays to zero for $p > p_c$. At $p_c \simeq 0.206$, $\varrho(t) \sim t^{\theta}$ grows as a power law with exponent $\theta \simeq 0.3175$, in good quantitative agreement with the critical exponent $\theta_{\text{DP}} \approx 0.3136$ governing the analogous growth of particle density in bond DP \cite{jensen_low-density_1999}. In the Supplemental Material \cite{SOM} we present additional numerical evidence for DP universality at the phase transition.

\textit{Decoding transition.---}The growth of the OTOC can be associated with the spreading of quantum information \cite{hosur_chaos_2016,yoshida_efficient_2017,xu_scrambling_2022}. However, this interpretation has important subtleties in our model. Suppose that an observer Alice stores a $k$-qubit message in the initial state of the system, which then undergoes time evolution by the circuit $U_t$. It may be tempting to associate the percolating OTOC to a capacity for another observer, Bob, to recover Alice's message from the qubits remaining in the system. However, closer examination shows that even when local operators develop large support on the system, they have substantially larger support on the qubits swapped out to the environment. As a result, Bob can never recover any fraction of Alice's message at long times. 

\begin{figure}[t]
	\includegraphics[width = \columnwidth]{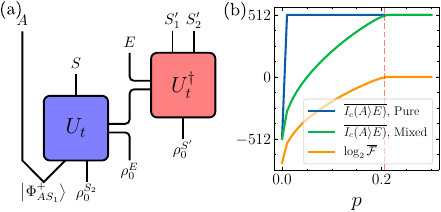}
	\caption{(a) Simple decoding protocol for recovering quantum information encoded initially within the system. Following random unitary evolution with swap gates on $SE$ via the circuit $U_t$ [Fig.~\ref{fig:fullwidth}(a)], swapped-out or ``radiated'' qubits are swapped back into a second system $S'$ via the reverse circuit $U_t^{\dagger}$. (b) Late-time log-averaged fidelity $\log_2 \overline{\mathcal{F}}$ (green) for the simple decoding protocol, and
	late-time average coherent information $\overline{I_c(A \rangle E)}$ for an initially maximally mixed environment (orange) and an initially pure environment (blue), as a function of swap rate $p$ in a system of size $N = 512$. Averages are taken over 400 Clifford circuit realizations. The dotted red line denotes $p_c \simeq 0.206$, as estimated from the OTOC.}
	\label{fig:3}
\end{figure}

To explicate the connection between operator spreading and the flow of quantum information in our model, we instead ask whether an eavesdropper Eve, who collects the qubits swapped out of the system, can recover Alice's message. We demonstrate here that the transition in operator spreading coincides with a transition in the fidelity with which Eve can decode Alice's quantum information from the radiated qubits using a simple decoding protocol, shown in Fig.~\ref{fig:3}(a) and discussed below. In the discussion, we comment on an analogy between our model and that of the Hayden-Preskill thought experiment \cite{hayden_black_2007}, and the relation between our decoding protocol and a similar protocol proposed for Hayden and Preskill's problem \cite{yoshida_efficient_2017}.

Alice's state is stored initially on the first $k$ qubits of $S$, denoted $S_1$. The remaining $N-k$ qubits of $S$ are denoted $S_2$. After evolving $SE$ by the circuit $U_t$, Eve collects the radiated qubits $E$ and attempts to decode Alice's state without access to $S$. To construct the decoder, Eve first introduces an extra set of $N$ qubits $S' = S_1' \cup S_2'$ initialized in an arbitrary state, then applies the reverse unitary circuit $U_t^{\dagger}$ on $S' E$ [Fig.~\ref{fig:3}(a)]. Deep in the nonpercolating phase, where Alice's state is always swapped entirely into the environment, such a decoding protocol will perfectly reproduce Alice's state on $S_1'$.

To quantify the success of Eve's decoding for general $p$, we encode Alice's state using a reference system $A$ initialized in a maximally entangled state $\ket*{\Phi^+_{A S_1}}$ with $S_1$, and compute the fidelity with the same state on $A S_1'$ following the decoding protocol:
\begin{equation}
\label{eq:F}
	\mathcal{F}(t) = \tr \qty{ \dyad{\Phi^+_{A S_1'}} U_t'^{\dagger} U_t \rho_0 U_t^{\dagger} U_t' } ,
\end{equation}
where $\rho_0 = \dyad*{\Phi^+_{AS_1}} \otimes \rho^{S_2 S' E}_0$ is the initial state on $ASS'E$, which consists of the entangled state $\dyad*{\Phi^+_{AS_1}}$ on $AS_1$ and an arbitrary product state $\rho_0^{S_2 S' E}$ on the remaining qubits, and $U_t'$ denotes the unitary circuit acting on $S'E$. A maximal fidelity $\mathcal{F} = 1$ implies that an arbitrary initial state on $S_1$ will be exactly reproduced on $S_1'$ at the end of the protocol, while a fidelity $\mathcal{F} = 2^{-2k}$ indicates that the final state on $S_1'$ is uncorrelated with the initial state of $S_1$.

For simplicity, we first assume that Alice's message contains $k=1$ qubit. Then, it is straightforward to show that in a given Clifford circuit, $\mathcal{F}(t)$ simply counts which of the operators $X_0(t)$, $Y_0(t)$, and $Z_0(t)$ are swapped entirely into the environment by time $t$. Through the mapping to the stochastic  model, this occurs for each such operator with probability $1-P_1(t)$, where $P_1(t)$ is the survival probability of the associated DP process initialized with a single particle at $t=0$. In the Supplemental Material \cite{SOM}, we show that this observation results in an average decoding fidelity $\overline{\mathcal{F}(t)} = 1 - \frac{3}{4} P_1(t)$ for $k = 1$ encoded qubit. More generally, for arbitrary $k$ we obtain the bounds
\begin{equation}
	1 - P_k(t)[1-2^{-2k}] \leq \overline{\mathcal{F}(t)} \leq 1 - P_1(t)[1-2^{-2k}] ,
\end{equation}
where $P_k(t)$ is the survival probability of $k$ initial particles arranged side-by-side. Notably, since $\mathcal{F}(t)$ is second-order in $U_t \otimes U_t^*$, this result holds identically in both the Haar and Clifford ensembles. In the nonpercolating phase $p>p_c$ each survival probability falls to zero exponentially quickly, resulting in a perfect decoding fidelity at late times as expected. On the other hand, for small $p < p_c$ the survival probability is large, and the decoding fidelity is close to that of a random final state on $AS'_1$. The numerical late-time behavior of $\log_2 \overline{\mathcal{F}}$ for $k = N$ is shown in Fig.~\ref{fig:3}(b).

\textit{Information transition.---}While we have shown that the decoding fidelity for the simple decoder of Fig. \ref{fig:3}(a) undergoes the same transition as the OTOC, we are more generally interested in the maximal amount of quantum information Eve can recover from the radiated qubits $E$ using \textit{any} decoding protocol. In other words, we would like to characterize the quantum channel capacity of the circuit $U_t$, regarded as a noisy quantum channel from $S_1$ to $E$ \cite{schumacher_sending_1996,schumacher_quantum_1996,lloyd_capacity_1997,divincenzo_quantum-channel_1998,barnum_information_1998,preskill_lecture_2018}. Towards this end we consider the coherent information $I_c(A \rangle E) = H_E - H_{AE}$ from $A$ to $E$, where $H_R = -\tr \rho^R_t \log \rho^R_t$ is the von Neumann entropy of subsystem $R$. The coherent information can then be used to lower-bound the single-shot quantum channel capacity of the circuit \cite{schumacher_sending_1996,lloyd_capacity_1997,devetak_private_2005,preskill_lecture_2018}. In this section it is useful to focus on the case $k = N$, although the generalization to arbitrary $k$ is straightforward \cite{SOM}.

Unlike the OTOC and the decoding fidelity given above, the behavior of $I_c(A \rangle E)$ depends strongly on the initial state of the qubits swapped into the system. First suppose that the swapped-in qubits are initialized in the maximally mixed state, $\rho^E_0 = \frac{1}{2^{N_E}} \mathds{1}$, where $N_E \sim pNt$ is the total number of swaps; physically, this corresponds to the case in which Eve has no prior knowledge of the qubits swapped into the circuit. Then, one can show diagrammatically \cite{SOM} that the subsystem purity $\tr \qty[ (\rho_t^{AE})^2 ] = 2^{N - N_E} \mathcal{F}$ is proportional to the decoding fidelity. This in turn implies that $I_c(A \rangle E) = N + \log_2 \mathcal{F}$ in individual Clifford circuit realizations. Although the logarithm prevents a simple statistical mechanics interpretation for the \textit{average} coherent information in either Clifford or Haar random circuits, we observe numerically [Fig.~\ref{fig:3}(b)] that $\overline{I_c(A \rangle E)}$ in the Clifford ensemble exhibits the same qualitative features as $N + \log_2 \overline{\mathcal{F}}$ and undergoes a transition at the same critical swap rate as the fidelity.

In contrast, suppose now that the swapped-in qubits are initialized in a definite pure state, $\rho^E_0 = \dyad{0}^{\otimes N_E}$. Physically, this case occurs when Eve has perfect knowledge of the initial state of each swapped-in qubit. Since the global state is now pure, we can compute $I_c(A \rangle E) = - I_c(A \rangle S)$ by tracing over the swapped-out qubits in $E$, upon which the swap operation becomes equivalent to an amplitude-damping channel \cite{nielsen_quantum_2010}. The system density matrix $\rho^S_t$ therefore evolves via a strictly contractive quantum channel with a unique fixed point, and rapidly forgets its initial conditions and approaches a unique steady state. As a result, we expect $\rho^{AS}_t \simeq \rho^A_t \otimes \rho^S_t$ to rapidly factorize into a product state. This implies that $I_c(A \rangle S) = -N$ after a finite timescale, indicating that no information can be transmitted from Alice to Bob through the system as expected.  But for a globally pure state, this immediately suggests $I_c(A \rangle E) = N$ is maximal. We confirm numerically [Fig.~\ref{fig:3}(b)] that in this case the coherent information is indeed maximal for any $p > 0$. Physically, this result implies that Eve can in principle use her knowledge of the swapped-in qubits to decode Alice's information from the radiated qubits for any $p > 0$ \cite{SOM}.

\textit{Discussion.---}We have demonstrated a DP phase transition in the operator dynamics and the flow of quantum information in a RUC which exchanges qubits with an environment. If an observer Alice stores a quantum message in the initial state of the system, another observer Eve can utilize a simple decoding protocol [Fig.~\ref{fig:3}(a)] to recover Alice's message from the radiated qubits with perfect fidelity in the nonpercolating phase $p > p_c$ and with imperfect fidelity in the percolating phase $p < p_c$.

The highly scrambling intrasystem unitary dynamics in our model plays an essential role in determining the critical swap rate $p_c$ and in obtaining a transition at a nonzero $p_c$. Indeed, in an alternate model consisting of non-scrambling dynamics such as free fermion evolution, it is straightforward to show that the non-percolating phase occurs for all $p>0$ \cite{SOM}. In contrast, the use of swap gates for system-environment interactions is not a physically crucial feature. For example, generic Haar-random gates between the system and environment can also drive a DP transition if the intrasystem dynamics is less scrambling, or if multiple rounds of system-environment interactions are allowed between layers of intrasystem gates.

We can draw an analogy between our model and the Hayden-Preskill thought experiment \cite{hayden_black_2007} by imagining our RUC as a black hole emitting Hawking radiation via qubit swaps with the environment. However, there are crucial differences: first, instead of assuming that the unitary circuit scrambles completely before emitting Hawking radiation, the radiation here is emitted dynamically throughout the scrambling process. Second, Eve does not have access to early radiation maximally entangled with the black hole prior to scrambling; as a result, Eve must collect a large amount (at least of order $N$) of radiation before she can decode Alice's message. Despite these differences, our model is a close analog to previous RUC models of evaporating black holes \cite{agarwal_toy_2020,piroli_random_2020}, and suggests the possibility of analogous phase transitions in the recoverability of quantum information in these models. Furthermore, while our decoding scheme [Fig.~\ref{fig:3}(a)] is closely analogous to the decoder proposed for Hayden and Preskill's problem \cite{yoshida_efficient_2017}, the decoding scheme in the present work does not require Grover search or postselection on exponentially rare measurement outcomes.

It is fruitful to compare the observed transition with the MIPT in monitored RUCs \cite{skinner_measurement-induced_2019,li_measurement-driven_2019,bao_theory_2020,jian_measurement-induced_2020}. In this setting, there is also a phase transition in the dynamics of quantum information: for small measurement rates below a threshold there is a finite capacity for Alice to transmit quantum information through the system, while above the threshold measurements are capable of destroying Alice's information \cite{choi_quantum_2020,gullans_dynamical_2020}. However, unlike the MIPT, the transition discussed in the present work requires no mid-circuit measurements and therefore does not suffer from a postselection problem. As a result, our transition does not face the same \textit{fundamental} barriers to experimental observation as the MIPT in monitored RUCs.

\begin{acknowledgments}
We thank Yimu Bao, Samuel Garratt, and Nishad Maskara for insightful discussions. We also thank David Huse for pointing out a subtle difference between the Haar and Clifford circuits in individual circuit realizations \cite{SOM}. E. A. is supported by the NSF QLCI program through Grant No. OMA-2016245 and by the U.S. Department of Energy, Office of Science, National Quantum Information Science Research Centers, Quantum Systems Accelerator (QSA). Z. W. is supported by the Berkeley Connect fellowship. This publication is funded in part by a QuantEmX grant from the Institute for Complex Adaptive Matter and the Gordon and Betty Moore Foundation through Grant GBMF9616 to Jamir Marino. S. P. K. and J. M. acknowledge support by the Dynamics and Topology Centre funded by the State of Rhineland Palatinate, and from the DFG through the TRR 288 - 422213477 (project B09) and TRR 306 Project-ID 429529648 (project D04). This research was done using services provided by the OSG Consortium \cite{osg07,osg09}, which is supported by the National Science Foundation awards \#2030508 and \#1836650.
\end{acknowledgments}

\bibliographystyle{apsrev4-2}
\bibliography{refs}

\end{document}


\title{Supplementary material for: Scrambling Transition in a Radiative Random Unitary Circuit}
\author{Zack Weinstein}
\thanks{ZW and SK contributed equally to this work.}
\affiliation{Department of Physics, University of California, Berkeley, CA 94720, USA}
\author{Shane P. Kelly}
\thanks{ZW and SK contributed equally to this work.}
\affiliation{Institute for Physics, Johannes Gutenberg University of Mainz, D-55099 Mainz, Germany}
\author{Jamir Marino}
\affiliation{Institute for Physics, Johannes Gutenberg University of Mainz, D-55099 Mainz, Germany}
\author{Ehud Altman}
\affiliation{Department of Physics, University of California, Berkeley, CA 94720, USA}
\affiliation{Materials Sciences Division, Lawrence Berkeley National Laboratory, Berkeley, CA 94720, USA}

\date{\today}
\maketitle

\tableofcontents

\section{OTOC Mapping for arbitrary qudit dimension}
\label{sec:OTOC_qudit}
In this section, we generalize our mapping of the OTOC dynamics to a directed percolation problem for arbitrary local Hilbert space dimension $q$. Unlike the $q = 2$ case of the main text where the Clifford ensemble can be leveraged for analytical simplicity, here we must directly perform the Haar average on the two-qudit unitary gates. We will see that this alternative approach reproduces the $q = 2$ results of the main text exactly. Our mapping slightly generalizes those of the works \cite{nahum_operator_2018,von_keyserlingk_operator_2018}. 

As in the main text, we consider a brick-wall Haar-random unitary circuit on qudits of local Hilbert space dimension $q$. In between layers of unitary gates, each system qudit is swapped with an environmental ancilla qudit with probability $p$. We are interested in the dynamics of the out-of-time-order correlator (OTOC) between two local Hermitian observables. For $q = 2$, the Pauli matrices formed a natural basis of Hermitian observables for a single qubit. For arbitrary $q$ we instead let $X^{a}$ ($a = 1, \ldots, q^2 - 1$) denote the set of $q^2 - 1$ generalized Gell-Mann matrices \cite{georgi_lie_1999}, traceless Hermitian matrices which generate the fundamental representation of SU$(q)$. They can always be chosen to satisfy the orthonormality condition $\tr[X^a X^b] = q \delta^{ab}$, which then implies the following useful identities\footnote{Any $q \times q$ matrix $M$ can be written as a linear combination of the $X^a$'s and the identity matrix $\mathds{1}$: $M = M^0 \mathds{1} + \sum_a M^a X^a$, where $M^0 = \frac{1}{q} \tr M$ and $M^a = \frac{1}{q} \tr[ M X^a ]$. Writing $M$ in terms of its matrix elements, we immediately derive the identity $\sum_a X^a_{ij} X^a_{k\ell} = q \delta_{i \ell} \delta_{jk} - \delta_{ij} \delta_{k \ell}$. The three quoted identities then immediately follow. The first of the three identities can also be derived immediately from Schur's lemma.}:
\begin{equation}
\label{eq:lie_identities}
	\sum_{a = 1}^{q^2 - 1} [X^a]^2 = (q^2 - 1) \mathds{1}_{q \times q}, \quad \sum_{a = 1}^{q^2 - 1} X^a X^b X^a = -X^b, \quad \sum_{a = 1}^{q^2 - 1} X^a X^b X^c X^a = q^2 \delta^{bc} \mathds{1}_{q \times q}- X^b X^c .
\end{equation}
Together with $X^0 \equiv \mathds{1}_{q \times q}$, we obtain a complete basis of Hermitian observables on a single qudit. In turn, these generate a basis of $q^{2M} - 1$ traceless Hermitian observables $X^{\vec{a}} = X^{a_1}_1 \ldots X^{a_M}_M$ on the many-body Hilbert space, where $M \equiv N + N_E$ is the total number of qudits in the system plus environment, plus the trivial operator string $X^{\vec{0}} = \mathds{1}_{q^M \times q^M}$. The orthogonality relation on $M$ qubits is generalized to $\tr[X^{\vec{a}} X^{\vec{b}}] = q^M \delta^{\vec{a} \vec{b}}$.

In analogy with the $q=2$ case, we define the OTOC as
\begin{equation}
\label{eq:otoc}
 	C_{ab}(x,t) = \frac{1}{q^2} \tr{ \rho_0 \comm{X^a_0(t)}{X^b_x}^{\dagger} \comm{X^a_0(t)}{X^b_x} } .
\end{equation}
To evaluate the ensemble average of $C_{ab}(x,t)$, we first expand $X^a_0(t) = \sum_{\vec{c}} w^a_{\vec{c}}(t) X^{\vec{c}}$ as a linear combination of operator strings $X^{\vec{c}}$ with weights $w^a_{\vec{c}}(t) = \frac{1}{q^M} \tr[ X^a_0(t) X^{\vec{c}}]$. By unitarity, these weights obey the normalization condition $\sum_{\vec{c}} [w^a_{\vec{c}}(t)]^2 = 1$. In the $q = 2$ Clifford circuit where $X^a_0(t)$ remains a Pauli string at all times, we have only one nonzero weight $w^a_{\vec{c}}(t) = \pm 1$ at each time step, and the ensemble average $\overline{[w^a_{\vec{c}}(t)]^2} = \overline{\abs{w^a_{\vec{c}}(t)}}$ is simply the probability for $X^a_0$ to evolve to the string $X^{\vec{c}}$. 

In the general $q > 2$ case, we cannot assume that $X^a_0(t)$ remains a single operator string at all times. Nevertheless, we obtain an effective Markovian dynamics upon averaging the weights $w^a_{\vec{c}}(t) w^a_{\vec{d}}(t)$, which can be written in terms of the average weights of the previous time step:
\begin{equation}
\label{eq:markov}
	\overline{w^a_{\vec{c}}(t+1) w^a_{\vec{d}}(t+1)} = \frac{1}{q^{2M}} \overline{ \tr \qty[ X^{\vec{c}} \mathcal{U}^{\dagger} X^a_0(t) \mathcal{U} ] \tr \qty[ X^{\vec{d}} \mathcal{U}^{\dagger} X^a_0(t) \mathcal{U} ] } = \frac{1}{q^{2M}} \sum_{\vec{e} \vec{g}} \overline{w^a_{\vec{e}}(t) w^a_{\vec{g}}(t)} \overline{ \tr[ X^{\vec{c}} \mathcal{U}^{\dagger} X^{\vec{e}} \mathcal{U} ] \tr[ X^{\vec{d}} \mathcal{U}^{\dagger} X^{\vec{g}} \mathcal{U} ] } .
\end{equation}
Here $\mathcal{U}$ denotes a single layer of unitary gates applied between timesteps $t$ and $t+1$, including both Haar-random unitaries on the system qudits and system-environment swaps occurring with probability $p$. Since the unitary gates at each time-step are uncorrelated, and $w^a_{\vec{e}}(t)$ depends only on unitary gates prior to the layer $\mathcal{U}$, the average factorizes as written above.

Since each trace in Eq.~(\ref{eq:markov}) additionally factorizes over each pair of qudits, it suffices to consider the average over a two-qudit system evolving via a $q^2 \times q^2$ Haar-random unitary gate $u$. The result from the two-qudit case can then immediately be applied to the remainder of the circuit. For the moment, set the swap rate $p = 0$ to focus on the Haar unitaries. Letting $X^{\vec{a}} = X^{a_1}_1 X^{a_2}_2$ denote a two-qudit operator string, the Haar average over the ensemble of two-qudit unitary gates $u$ is given via the usual Weingarten calculus \cite{collins_moments_2003,collins_integration_2006,nahum_quantum_2017,nahum_operator_2018} as
\begin{equation}
	\begin{split}
		&\overline{\tr \qty[ X^{\vec{c}} u^{\dagger} X^{\vec{e}} u ]  \tr \qty[ X^{\vec{d}} u^{\dagger} X^{\vec{g}} u ]} \\
		& \quad \quad  = \frac{1}{q^4 - 1} \qty{ \tr \qty[ X^{\vec{c}} X^{\vec{d}} ] \tr[ X^{\vec{e}} X^{\vec{g}} ] + \tr X^{\vec{c}} \tr X^{\vec{e}} \tr X^{\vec{d}} \tr X^{\vec{g}} - \frac{1}{q^2} \qty( \tr[X^{\vec{c}} X^{\vec{d}}] \tr X^{\vec{e}} \tr X^{\vec{g}} + \tr X^{\vec{c}} \tr X^{\vec{d}} \tr[ X^{\vec{e}} X^{\vec{g}}] ) } .
	\end{split}
\end{equation}
The above expression vanishes unless $\vec{c} = \vec{d}$ and $\vec{e} = \vec{g}$, and unless $\vec{e} = \vec{g} = \vec{0}$ whenever $\vec{c} = \vec{d} = \vec{0}$. There are therefore only two cases: the trivial  string $X^{\vec{e}} = X^{\vec{g}} = \mathds{1}$ evolves to the trivial operator string $X^{\vec{c}} = X^{\vec{d}} = \mathds{1}_{q^2 \times q^2}$ with unit probability, while any nontrivial operator string evolves evolves to each of the $q^4 - 1$ nontrivial operator strings with equal weight. Explicitly, Eq.~(\ref{eq:markov}) for a two-qudit system reads
\begin{equation}
	\overline{w^a_{\vec{c}}(t+1) w^a_{\vec{d}}(t+1)} = \delta_{\vec{c}, \vec{d}} \qty( [w^a_{\vec{0}}(t)]^2 \delta_{\vec{c}, \vec{0}} + \frac{(1 - \delta_{\vec{c},\vec{0}})}{q^4 - 1}\sum_{\vec{e} \neq \vec{0}} [w^a_{\vec{e}}(t)]^2 ) \quad (p = 0) .
\end{equation}
Extending this two-qudit result to the full system of $N/2$ unitary gates at each time step, we can interpret the evolution of $\overline{[w^a_{\vec{c}}(t)]^2}$ as the Markovian evolution of a probability distribution of particles on a tilted square lattice. Vertices of the lattice reside at the center of each unitary gate, and $q^2 - 1$ species of particles travel upward through the lattice. If at least one particle enters a vertex from below, either one or two particles can exit the top of the vertex; thus, in the case $p = 0$, we obtain a directed percolation-like model in which particles can either diffuse, spread, or coagulate, but never annihilate. The various probabilities for the propagation of particles from an occupied vertex are obtained simply by counting two-qudit operators: there are $q^2 - 1$ operators $X^{\vec{a}} = X^{a_1}_1 X^0_2$ with support on only the first qudit (and similarly for the second qudit), and $(q^2 - 1)^2$ operators with support on both qudits. In the absence of swaps, we therefore obtain the following stochastic update rules for an occupied vertex:

\begin{equation}
	\tikz{
		\draw[line width = 1.0pt] (0,0) -- (-0.5, 0.5);
		\draw[line width = 1.0pt] (0,0) -- (0.5, 0.5);
		\filldraw (0,0) circle (0.05);
		\node at (0, -0.5) {$ \frac{(q^2-1)^2}{q^4-1} $,};
	} 
        \quad \tikz{
		\draw[line width = 1.0pt] (0,0) -- (-0.5, 0.5);
		\draw[dotted, line width = 1.0pt] (0,0) -- (0.5, 0.5);
		\filldraw (0,0) circle (0.05);
		\node at (0, -0.5) {$ \frac{q^2 - 1}{q^4 - 1} $,};
	}
        \quad \tikz{
		\draw[dotted, line width = 1.0pt] (0, 0) -- (-0.5, 0.5);
		\draw[line width = 1.0pt] (0, 0) -- (0.5, 0.5);
		\filldraw(0, 0) circle (0.05);
		\node at (0, -0.5) {$ \frac{q^2 - 1}{q^4 - 1} $,};
	} 
        \quad \tikz{
		\draw[dotted, line width = 1.0pt] (0,0) -- (-0.5, 0.5);
		\draw[dotted, line width = 1.0pt] (0,0) -- (0.5, 0.5);
		\filldraw (0,0) circle (0.05);
		\node at (0, -0.5) {$ 0 $};
		\node at (2, 0) {$(p = 0) ,$};
	}
\end{equation}
where a dark line indicates the propagation of a particle, while a dotted line indicates no propagation of a particle. 

Upon introducing a finite rate of swaps $p > 0$ following each layer of unitary gates, the operator content in  each string $X^{\vec{c}}$ in the system is traded for the trivial operator content of each swapped-in qudit. In this way, particles effectively hop from the system to the environment. The new stochastic update rules are obtained from the $p = 0$ rules above simply by restricting propagation along each edge with probability $p$:
\begin{equation}
\label{eq:branching_rules}
	\tikz{
		\draw[line width = 1.0pt] (0,0) -- (-0.5, 0.5);
		\draw[line width = 1.0pt] (0,0) -- (0.5, 0.5);
		\filldraw (0,0) circle (0.05);
		\node at (0, -0.5) {$ \frac{(q^2-1)^2}{q^4-1} (1-p)^2 $,};
	}
        \quad \tikz{
		\draw[line width = 1.0pt] (0,0) -- (-0.5, 0.5);
		\draw[dotted, line width = 1.0pt] (0,0) -- (0.5, 0.5);
		\filldraw (0,0) circle (0.05);
		\node at (0, -0.5) {$ \frac{q^2 - 1}{q^4 - 1}(1-p) + \frac{(q^2-1)^2}{q^4-1} p(1-p) $,};
	}
        \quad \tikz{
		\draw[dotted, line width = 1.0pt] (0, 0) -- (-0.5, 0.5);
		\draw[line width = 1.0pt] (0, 0) -- (0.5, 0.5);
		\filldraw(0, 0) circle (0.05);
		\node at (0, -0.5) {$ \frac{q^2 - 1}{q^4 - 1}(1-p) + \frac{(q^2-1)^2}{q^4-1} p(1-p) $,};
	}
        \quad \tikz{
		\draw[dotted, line width = 1.0pt] (0,0) -- (-0.5, 0.5);
		\draw[dotted, line width = 1.0pt] (0,0) -- (0.5, 0.5);
		\filldraw (0,0) circle (0.05);
		\node at (0, -0.5) {$ \frac{(q^2-1)^2}{q^4-1} p^2 + 2p\frac{q^2 - 1}{q^4 - 1} $.};
	}
\end{equation}
It is useful to note that in the $q \to \infty$ limit, each edge is open with probability $(1-p)$ and closed with probability $p$, yielding exactly bond-directed percolation. For any finite $q$, the two edges are not perfectly independent. Nevertheless, even for $q = 2$ the corrections to bond-directed percolation are small, and we expect that the stochastic process will undergo a phase transition in the directed percolation universality class.

Returning to the OTOC, we have on average
\begin{equation}
\label{eq:avg_otoc}
	\overline{C_{ab}(x,t)} = \frac{1}{q^2} \sum_{\vec{c}} \overline{[w^a_{\vec{c}}(t)]^2}  \tr \qty{ \rho_0 \comm{X^{\vec{c}}}{X^b_x}^{\dagger} \comm{X^{\vec{c}}}{X^b_x} } .
\end{equation}
Here we have expanded both instances of $X^a_0(t)$ using the weights $w^a_{\vec{c}}$ and applied the Haar average. We can simplify this expression further by noting that the value of each squared weight $\overline{[w^a_{\vec{c}}(t)]^2}$ corresponding to a particular distribution of particles is independent of the particle species on each occupied site, since each particle species evolves to each other particle species with equal weight. Let $n_j(t) = 0,1$ denote the occupation of a particular site $j$ at time $t$ according to the distribution of particles specified by $\vec{c}$, with total particle number $N(t) = \sum_j n_j(t)$. Furthermore, let $\mathcal{P}( \qty{n_j(t)} )$ denote the probability for a particular distribution of these occupation numbers, regardless of the particle species. Then, 
\begin{equation}
	\overline{[w^a_{\vec{c}}(t)]^2} = \frac{1}{(q^2 - 1)^{N(t)}} \mathcal{P}( \qty{n_j(t)}) .
\end{equation}
The factor $(q^2 - 1)^{N(t)}$ in the denominator is the total number of weights $\overline{[w^a_{\vec{c}}(t)]^2}$ with the same particle distribution $\qty{n_j(t)}$. We can then reorganize the sum over operator strings in Eq.~(\ref{eq:avg_otoc}) into a sum over occupation numbers:
\begin{equation}
	\overline{C_{ab}(x,t)} = \frac{1}{q^2} \sum_{\qty{ n_j(t) }} \frac{\mathcal{P}(\qty{n_j(t)})}{(q^2 - 1)^{N(t)}} \sum_{\vec{c}}' \tr \qty{ \rho_0 \comm{X^{\vec{c}}}{X^b_x}^{\dagger} \comm{X^{\vec{c}}}{X^b_x} } .
\end{equation}
Here the first sum includes all configurations of occupation numbers, while the latter primed sum includes only those operator strings $\vec{c}$ consistent with the occupation numbers $\qty{n_j(t)}$. For each occupied site $j$ away from site $x$, the sum over all nontrivial operators $\sum_{c_j = 1}^{q^2 - 1} [X^{c_j}_j]^2$ simply yields a factor of $q^2 - 1$ via the first identity of Eq. (\ref{eq:lie_identities}). At site $x$, there are two possibilities: if $n_x(t) = 0$, then the primed sum requires $c_x = 0$ and the entire expression vanishes. If $n_x(t) = 1$, then the primed sum yields
\begin{equation}
	\begin{split}
		\sum_{c_x = 1}^{q^2 - 1} \comm{X^{c_x}_x}{X^b_x}^{\dagger} \comm{X^{c_x}_x}{X^b_x} &= \sum_{c_x = 1}^{q^2 - 1} \qty{ X^b_x X^{c_x}_x X^{c_x}_x X^b_x - X^{c_x}_x X^b_x X^{c_x}_x X^b_x - X^b_x X^{c_x}_x X^b_x X^{c_x}_x + X^{c_x}_x X^b_x X^b_x X^{c_x}_x } \\
		&= q^2 \qty{ \mathds{1} + [X^b_x]^2 } ,
	\end{split}
\end{equation}
where we have again used the identities (\ref{eq:lie_identities}). Combining these elements, we finally arrive at the remarkably simple expression:
\begin{equation}
	\overline{C_{ab}(x,t)} = \frac{1 + \tr \qty{ \rho_0 [X^b_x]^2 }}{q^2 - 1} \overline{n_x(t)} ,
\end{equation}
where $\overline{n_x(t)} = \sum_{\qty{n_j(t)}:n_x(t)=1} \mathcal{P}(\qty{n_j(t)})$ is the probability of finding a particle at site $x$ at time $t$. Up to a static prefactor, the average OTOC is simply given by the total probability for a particle to be present at position $x$ at time $t$. It is interesting to note that all dependence on the initial state $\rho_0$ and the probe operator $X^b_x$ appears only in the prefactor, which has no effect on the universal behavior of the OTOC. In the case $q = 2$ the $X^a$ operators are the Pauli matrices and square to one, giving an OTOC $\overline{C_{ab}(x,t)} = \frac{2}{3} \overline{n_x(t)}$ as in the main text.

As a final remark, we comment on a subtle difference between the Haar and Clifford circuits in the $q = 2$ case\footnote{We thank David Huse for pointing out this subtlety to us.}. Since the Clifford ensemble forms a unitary three-design \cite{webb_clifford_2016,zhu_multiqubit_2017}, the average behavior of the OTOC is identical between the Haar and Clifford circuits. However, there are crucial differences in individual circuit realizations. In the Clifford circuit, the evolution of $X_0(t)$ exactly realizes the effective stochastic model, as discussed in the main text. As a result, $X_0(t)$ is guaranteed to vanish uniformly after a finite time in almost all circuit realizations above the critical point. In contrast, in the Haar circuit, $X_0(t)$ does not realize the effective stochastic model in individual circuit realizations: instead, it is a sum of exponentially many nonzero weights $w_{\vec{c}}(t) = \frac{1}{2^M} \tr[X_0(t) X^{\vec{c}}]$, where $X^{\vec{c}}$ is a string of Pauli operators. As a result, fluctuations in the OTOC about its mean in a given circuit realization are strongly suppressed compared to the Clifford circuit \cite{nahum_operator_2018,von_keyserlingk_operator_2018}. In particular, the OTOC in a given circuit realization continues to propagate within the circuit past the critical point $p_c \simeq 0.206$, albeit with an exponential suppression in time, as in the ensemble-averaged OTOC. In principle, there is a second phase transition at $p_{c2} \approx 0.355$, corresponding to the DP phase transition in the underlying lattice itself (rather than in the effective model), above which the OTOC will vanish uniformly at a finite time in almost all Haar circuit realizations. This transition is likely difficult to observe in practice, since the OTOC is anyway exponentially suppressed in time.

\section{Estimation of light cone velocity}
For the particular case $p = 0$, Refs.~\cite{nahum_operator_2018,von_keyserlingk_operator_2018} showed that the OTOC spreads along a light cone with velocity $v_B = (q^2 - 1) / (q^2 + 1)$, with a light-cone front which broadens as $\sqrt{t}$. As in ordinary bond-DP, we expect an increasing swap rate $p$ to result in a narrowing of the light cone, and correspondingly a decrease of the light cone velocity. In this section we give a ``mean-field'' calculation of the light cone velocity, which we expect to be valid for small nonzero swap rates.

We first estimate the density of particles within the bulk of the light cone by assuming the bulk occupation reaches a steady-state of fixed particle density with no nontrivial correlations between sites. Here, we must distinguish between the \textit{edge} density $\varrho_e$ and the $\textit{vertex}$ density $\varrho_v$; the former governs the probability for a given edge to be occupied, while the latter governs the probability for a given vertex in the tilted square lattice to be occupied. Assuming $\varrho_e$ and $\varrho_v$ are constant in time and space, and using the branching rules (\ref{eq:branching_rules}), we can derive the following two relations between the two densities:
\begin{equation}
	\varrho_e = (1-p)\qty[ \frac{q^2}{q^2 + 1} ] \varrho_v, \quad \varrho_v = 2 \varrho_e - \varrho_e^2 .
\end{equation}
The former expression simply says that the probability for an edge to be occupied is equal to the probability that the vertex below it is occupied, times the probability that a particle propagates along the edge. The latter expression says that the probability for a vertex to be occupied is equal to the probability that at least one of the two edges feeding into the vertex is occupied. From these, we can immediately estimate the steady-state densities\footnote{As an aside, note that these equations can be used to predict a mean-field transition point at $p = \frac{1}{2}(1 - q^{-2})$, or $\frac{3}{8}$ for $q = 2$. As might be expected on physical grounds, this mean-field analysis over-estimates the transition point.}:
\begin{equation}
	\varrho_e = \frac{q^2 - 1 - 2pq^2}{(1-p)q^2} = \frac{q^2 - 1}{q^2} - \frac{q^2 + 1}{q^2}p + \ldots, \quad \varrho_v  = \frac{(q^2 + 1)(q^2 - 1 - 2pq^2)}{(1-p)^2 q^4} = \frac{q^4 - 1}{q^4} - 2p \frac{q^2 + 1}{q^4} + \ldots ,
\end{equation}
where the ellipses denotes higher-order contributions in $p$. Note that the particular case $q = 2$ and $p = 0$ estimates $\varrho_e = \frac{3}{4}$, resulting in an OTOC $\overline{C(x,t)} = \frac{1}{2}$.

Using these expressions, we can estimate the light cone velocity for small $p$ as follows. Consider the trajectory of the rightmost particle in the system, whose motion is governed by the branching rules (\ref{eq:branching_rules}). As the particle leaves a given vertex, there are three possibilities: the particle can hop one vertex to the right, or the particle can hop one vertex to the left, or it can die along both edges leaving the vertex. These occur with probabilities $P_r$, $P_{\ell}$, and $P_d$ respectively, given by
\begin{equation}
	P_r = (1-p) \frac{q^2}{q^2 + 1}, \quad P_{\ell} = (1-p)\frac{1}{q^2 + 1} + p(1-p)\frac{q^2 - 1}{q^2 + 1}, \quad P_d = p^2 \frac{q^2 - 1}{q^2 + 1} + 2p \frac{1}{q^2 + 1} .
\end{equation}
If the rightmost particle dies, then the next rightmost occupied vertex is on average a distance roughly $2/\varrho_v$ to the left\footnote{The factor of two arises because the vertices of the tilted square lattice sit two lattice spacings apart.}. The behavior of the rightmost particle therefore undergoes a biased random walk with three possible steps. The average distance traveled in each time step gives the light cone velocity:
\begin{equation}
   \begin{split}
    v_B \simeq P_r - P_{\ell} - \frac{2}{\rho_v} P_d &= \frac{(1-p)^2 \left((2 p (p+1)-1) q^6+(1-2 (p-2) p) q^4+(1-2 p)
   q^2-1\right)}{\left(q^2+1\right)^2 \left((2 p-1) q^2+1\right)} \\
   &= \frac{q^2-1}{q^2+1} -\frac{2 p \left(q^6+q^4-q^2+1\right)}{\left(q^2-1\right)
   \left(q^2+1\right)^2}+\ldots ,
   \end{split}
\end{equation}
where the ellipsis in the last expression denotes higher-order corrections in $p$. Since the rightmost particle continues to undergo a random walk for nonzero $p$, we also predict that the $\sqrt{t}$ broadening of the light cone front persists for nonzero swap rates.

In this derivation, we have assumed that the steady-state probability distribution has no correlations in the occupations of nearby vertices or edges, and we have assumed that the average distance between particles is equal to the inverse of the density. Both of these expectations are reasonable away from the transition, when the local density has small fluctuations about their mean values. Near the transition, we instead expect the light cone to be set by the ratio of the correlation length $\xi_{\perp}$ to the correlation time $\xi_{\parallel}$, which vanishes near the critical point as
\begin{equation}
    v_B \simeq \frac{\xi_{\perp}}{\xi_{\parallel}} \sim \frac{|p-p_c|^{-\nu_{\perp}}}{|p-p_c|^{-\nu_{\parallel}}} = |p-p_c|^{\nu_{\perp}(z-1)}
\end{equation}
where $z = \nu_{\parallel} / \nu_{\perp}$ is the dynamical exponent. For directed percolation $z \approx 1.581$ and $\nu_{\perp} \approx 1.097$ \cite{jensen_low-density_1999}, so that we expect the light cone velocity to vanish as $|p-p_c|^{0.637}$ near the critical point.

\section{Additional numerical details}
In this section, we provide additional details of our numerical simulations and demonstrate directed percolation (DP) universality at the transition by determining the critical exponents. We also discuss the simulation of information-theoretic quantities such as the coherent information, which requires some modification of the numerical approach used to compute the OTOC.

\subsection{Clifford Simulation}
We perform our Clifford simulations of the circuit using the stabilizer formalism; see Refs. \cite{gottesman_heisenberg_1998,aaronson_improved_2004,nahum_operator_2018,li_measurement-driven_2019} for a detailed discussion. To compute the OTOC we need only to track the evolution of a single stabilizer $X_0(t)$, which remains a Pauli string at all times under Clifford unitary evolution. Using the normalization convention of Eq.~(\ref{eq:otoc}) with $q = 2$, the value of the OTOC $C(x,t)$ equals one whenever $X_0(t)$ has Pauli content $X$ or $Z$ at site $x$, and is zero otherwise. We need not explicitly keep track of the Pauli content on environment qubits; instead, we simply set the Pauli content at a given site to zero whenever a swap occurs at that location. 



\subsection{DP Universality, Critical Exponents}

\begin{figure}[h]
	\centering
	\includegraphics[width=0.7\textwidth]{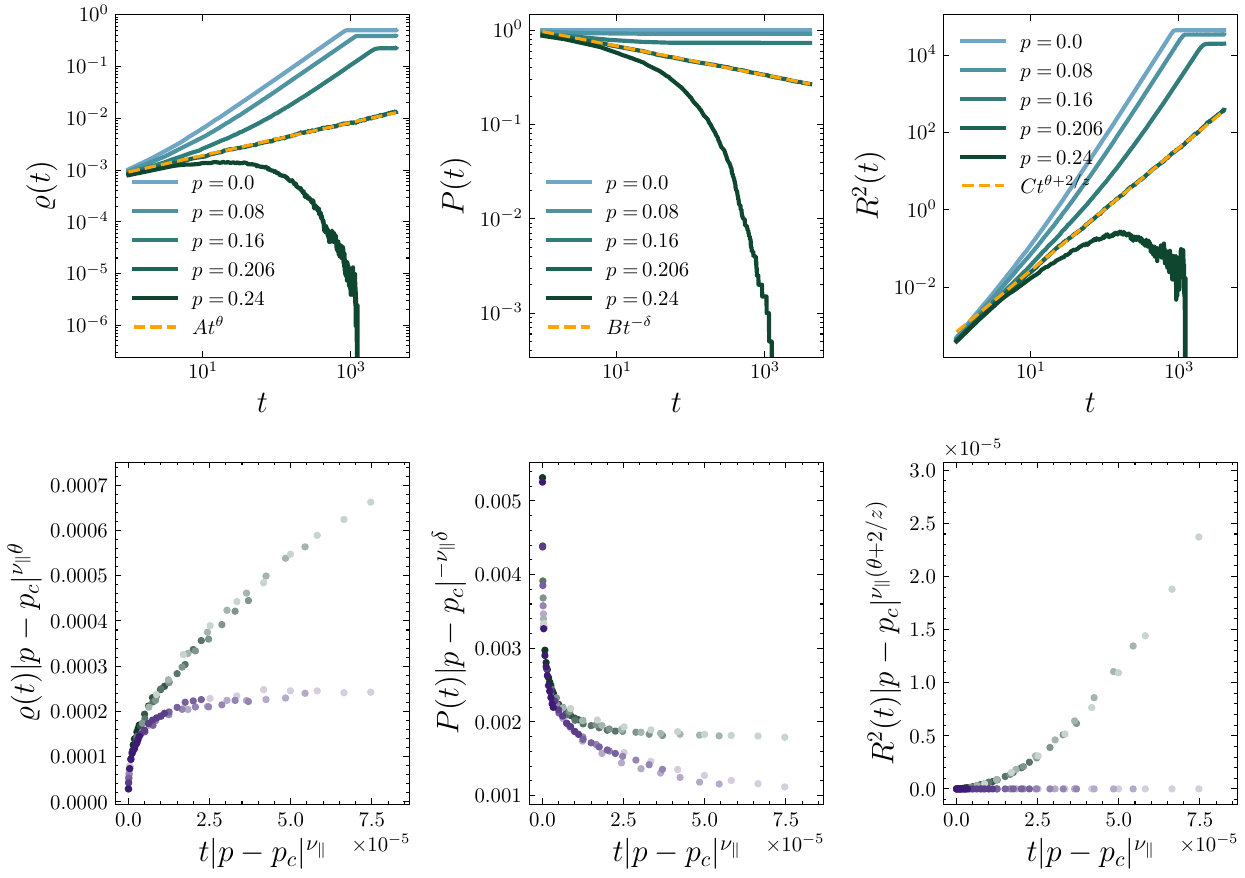}
	\caption{Top: Particle density $\varrho(t)$, survival probability $P(t)$, and mean-squared spreading $R^2(t)$ as a function of time on log-log scales, for various swap rates. Dotted orange line depicts best power-law curve fit at the critical point $p_c \simeq 0.206$. Bottom: finite-size scaling collapses for the same three observables, for swap rates $p \in \qty{0.203, 0.2035, 0.204, 0.2045, 0.205, 0.2055}$ below the critical point (green), and for swap rates $p \in \qty{0.2065, 0.207, 0.2075, 0.208, 0.2085, 0.209}$ above the critical point (purple). Averages are taken over 2,000 circuit realizations.}
	\label{fig:PNR}
\end{figure}

Since the average OTOC $\overline{C(x,t)}$ measures the density of particles at site $x$ and time $t$ starting from a single initial particle at $(x,t) = (0,0)$, we identify the OTOC with the pair-connectedness function of the DP-like spreading process \cite{hinrichsen_nonequilibrium_2000}. From this identification, we can define three prototypical numerical probes of directed percolation: the particle density $\varrho(t)$, survival probability $P(t)$, and mean-squared spreading $R^2(t)$. In terms of the OTOC, these are defined here as
\begin{equation}
	\varrho(t) = \frac{1}{N} \sum_x \overline{C(x,t)}, \quad P(t) = \overline{ \Theta \qty(\sum_x C(x,t) > 0) }, \quad R^2(t) = \frac{1}{N} \sum_x x^2 \overline{C(x,t)} .
\end{equation}
In the percolating phase $p < p_c$, each of these quantities saturates at a finite value. In the nonpercolating phase $p > p_c$, each of these quantities vanishes after an $\mathcal{O}(1)$ time independent of system size. At the critical point, we expect these quantities to exhibit power-law scaling determined by the three critical exponents $\Theta$, $\delta$, and $z$, valid for times large compared to the microscopic timescale but small compared to $N^z$ \cite{hinrichsen_nonequilibrium_2000}:
\begin{equation}
	\varrho(t) \sim t^{\Theta}, \quad P(t) \sim t^{-\delta}, \quad R^2(t) \sim t^{\Theta + 2/z}
\end{equation}
These basic features of the two phases, and the power-law scaling at the critical point, are demonstrated in Fig.~\ref{fig:PNR}. 

We determine the critical point $p_c \simeq 0.206$ from the best linear fit of $\varrho(t)$ on a log-log scale. The slope of the linear fit then determines $\Theta \simeq 0.3175$. The corresponding slopes for $P(t)$ and $R^2(t)$ at the critical point similarly determine $\delta \simeq 0.1537$ and $z \simeq 1.630$. Each of these quantities compares favorably with the accepted DP values \cite{jensen_low-density_1999}, as shown in table~\ref{tab:exponents}.

\begin{table}[h]
\centering
\begin{tabular}{l|lll}
         & OTOC & DP & Error \\ \cline{2-4} 
$\Theta$ & 0.3175   & 0.3136   & 1.2\% \\
$\delta$ & 0.1537   & 0.1595   & 3.6\% \\
$z$      & 1.630    & 1.581    & 3.1\%
\end{tabular}
\caption{Critical exponents $\Theta$, $\delta$, and $z$ obtained numerically from the OTOC, compared to the corresponding accepted critical exponents of directed percolation \cite{jensen_low-density_1999}.}
\label{tab:exponents}
\end{table}

More generally, in the vicinity of the critical point, each of these quantities are expected to exhibit the following scaling forms \cite{hinrichsen_nonequilibrium_2000}:
\begin{equation}
\label{eq:scaling_functions}
	\begin{split}
		\varrho(t) &= t^{\Theta} \Phi_{\varrho}^{\pm}(t |p-p_c|^{\nu_{\parallel}}, tN^{-z}), \quad P(t) = t^{-\delta} \Phi^{\pm}_P(t|p-p_c|^{\nu_{\parallel}}, t N^{-z}), \quad R^2(t) = t^{\Theta + 2/z} \Phi_R^{\pm}(t |p-p_c|^{\nu_{\parallel}}, tN^{-z}) ,
	\end{split}
\end{equation}
where we have defined six universal scaling functions $\Phi^{\pm}$ for the three quantities above and below the critical point, as well as an additional critical exponent $\nu_{\parallel}$ governing the correlation length in the time direction. In Fig.~\ref{fig:PNR}, we demonstrate that a handful of swap rates immediately above and below the critical point collaspe onto these universal functions, using the accepted DP critical exponents \cite{jensen_low-density_1999}. Note that each vertical axis is scaled as $|p-p_c|^{y \nu_{\parallel}}$, rather than against $t^{-y}$, resulting in slightly different (but equivalent) scaling functions compared to Eq.~(\ref{eq:scaling_functions}).

\begin{figure}[h]
	\centering
	\includegraphics[width = 0.6\textwidth]{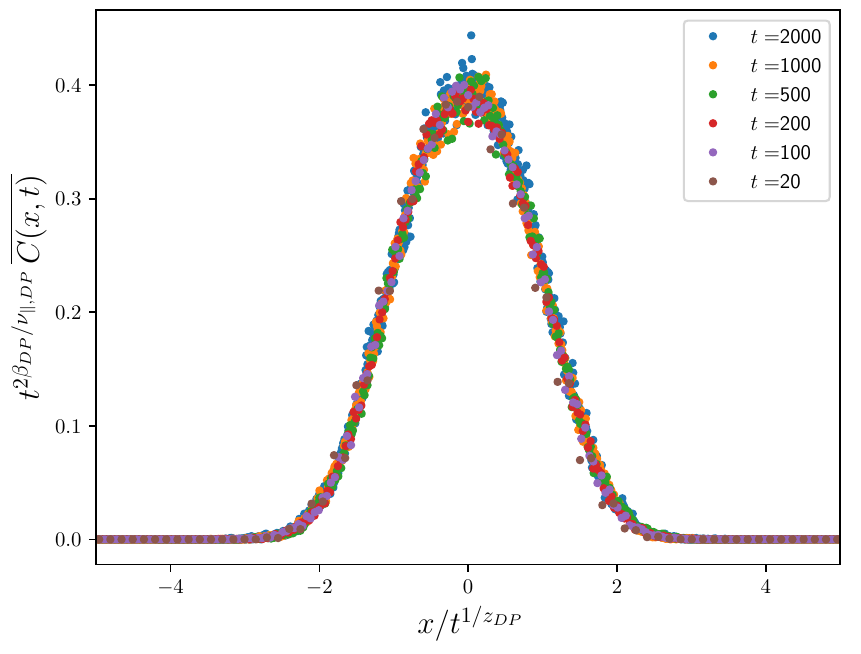}
	\caption{Scaling collapse of the averaged OTOC $\overline{C(x,t)}$ at the estimated critical swap rate $p = 0.206$, for various time slices $t$. Axes are scaled according to the scaling function (\ref{eq:otoc_scaling_fn}), using the accepted DP critical exponents \cite{jensen_low-density_1999}.}
	\label{fig:otoc_collapse}
\end{figure}

As a final probe of DP universality, we note that the OTOC itself is expected to obey the following scaling form \cite{hinrichsen_nonequilibrium_2000}
\begin{equation}
\label{eq:otoc_scaling_fn}
	\overline{C(x,t)} = t^{-(\beta + \beta')/\nu_{\parallel}} F(xt^{-1/z}, t |p-p_c|^{\nu_{\parallel}}, tN^{-z}) .
\end{equation}
In Fig.~\ref{fig:otoc_collapse} we plot the spatial profile of the OTOC averaged over many circuit realizations at the critical point, for several different times, and with axes scaled according to the accepted DP exponents \cite{jensen_low-density_1999}. As expected, we find that each curve collapses onto a universal scaling function, providing strong evidence for DP universality.

\subsection{Information Dynamics}
While numerically computing the OTOC required tracking only a single stabilizer evolving under random Clifford evolution, computing the dynamics of information-theoretic quantities such as the coherent information is more involved. We consider an initial stabilizer state specified by a list of $d$ generators $\mathcal{G} = \qty{g_i}_{i = 1}^d$, a set of independent and commuting Pauli strings. These generators form a basis for a stabilizer group $\mathcal{S} = \expval{g_1, \ldots , g_d}$, an abelian group of commuting Pauli strings. The density matrix is initially given by
\begin{equation}
	\rho_0 = \frac{2^d}{2^M} \prod_{i = 1}^d \qty( \frac{1 + g_i}{2} ) ,
\end{equation}
where $M$ is the total number of qubits. By tracking the evolution of the generators $g_i(t) = U_t g_i U_t^{\dagger}$, which remain Pauli strings at all times under Clifford evolution, we can efficiently simulate the time evolution of $\rho_t = U_t \rho_0 U_t^{\dagger}$. A comprehensive introduction to Clifford simulation can be found in Refs.~\cite{gottesman_heisenberg_1998,aaronson_improved_2004,nahum_operator_2018,li_measurement-driven_2019}; our goal here is to explain how information-theoretic quantities can be computed in our particular model featuring $N_E \sim pLt$ environment qubits (which cannot always be traced over) without explicitly keeping track of each generator's Pauli content on $E$.

We initialize the first $k$ qubits of the system, denoted $S_1$, in a Bell pair with a reference system $A$. This Bell pair gives rise to $2k$ initial stabilizer generators $X_{A,j} X_j$ and $Z_{A,j} Z_j$ for $1 \leq j \leq k$. For the remaining $N-k$ qubits of the system (denoted $S_2$) and the $N_E$ environment qubits, we consider three cases: first, $(i)$ we suppose $S_2$ and $E$ are initialized in the maximally mixed state; second, $(ii)$ we suppose $S_2$ is initialized in the pure product state $\dyad{0}^{\otimes N-k}$ while $E$ remains maximally mixed; and finally, $(iii)$ we suppose both $S_2$ and $E$ are initialized in the pure product state $\dyad{0}^{\otimes N-k+N_E}$. We first focus on the former two cases, in which the number of generators $d$ remains fixed in time independent of the size of $E$.

To simulate a system-environment swap at site $j$, it is insufficient to simply eliminate the Pauli content at site $j$ for each generator. Instead, viewing $\mathcal{S}$ as a $d$-dimensional vector space over the finite field $\mathbb{F}_2 = \qty{0,1}$ with basis vectors $\mathcal{G}$, it is useful to first perform a change of basis on the stabilizer group such that $\mathcal{G} = \mathcal{G}_{AS} \cup \mathcal{G}_{AS}^{\perp}$ at all times, where $\mathcal{G}_{AS}$ is a basis for the subspace $\mathcal{S}_{AS} \subset \mathcal{S}$ of stabilizers with trivial Pauli content outside of $AS$, while $\mathcal{G}_{AS}^{\perp}$ is a basis for the orthogonal complement of $\mathcal{S}_{AS}$. Explicitly, if a swap then occurs at site $j$, we perform a change of basis on $\mathcal{G}_{AS}$ such that at most one generator has Pauli-$X$ content on site $j$, and at most one generator has Pauli-$Z$ content on site $j$. We then move each such generator with support on site $j$ from $\mathcal{G}_{AS}$ to $\mathcal{G}_{AS}^{\perp}$, and eliminate the Pauli content in all generators on site $j$ to simulate the swap with an untracked environment qubit. In this way, we keep track of the Pauli content on $AS$ for the full set of generators, and we can obtain the generators of the subgroup $\mathcal{S}_{AS}$ at any time by simply discarding the generators $\mathcal{G}_{AS}^{\perp}$, without ever explicitly tracking any generator's Pauli content on $E$. 

We will be interested in computing the coherent information $I_c(A \rangle E) = H_E - H_{AE}$ from the reference into the environment, as well as the coherent information $I_c(A \rangle S) = H_S - H_{AS}$ into the system. Both quantities stem from the von Neumann entropy $H_R = -\tr \rho^R_t \log \rho^R_t$ of subsystem $R$, which can be computed efficiently from the list of generators via \cite{hamma_bipartite_2005,hamma_ground_2005,nahum_quantum_2017,li_measurement-driven_2019}
\begin{equation}
\label{eq:EE_stab_1}
    H_R = N_R - \abs{\mathcal{G}_R} ,
\end{equation}
where $N_R$ is the number of qubits in subsystem $R$, and $\abs{\mathcal{G}_R}$ is the number of generators for the subgroup $\mathcal{S}_R \subset \mathcal{S}$ of Pauli strings with no nontrivial support outside the region $R$. Once again viewing $\mathcal{S}$ as a vector space over $\mathbb{F}_2$, we can write $\abs{\mathcal{G}_R} = \dim \ker \text{proj}_{R^c} (\mathcal{S})$, where $\text{proj}_{R^c}$ is a linear operator which eliminates the Pauli content of a string $g$ within subsystem $R$. Using the rank-nullity theorem of linear algebra \cite{hoffman1971linear}, we can then express $H_R$ as
\begin{equation}
\label{eq:EE_stab_2}
	H_R = N_R - d + \rank \text{proj}_{R^c} (\mathcal{S}) .
\end{equation}
The latter expression can be determined immediately from the list of generators $\mathcal{G}$ using Gaussian elimination \cite{nahum_quantum_2017,li_measurement-driven_2019}. 

In the case of a maximally mixed environment $E$ (cases $(i)$ and $(ii)$), Eq.~(\ref{eq:EE_stab_2}) is sufficient for computing the quantities $H_E$, $H_{SE}$, and $H_{AE}$, since $\rank \text{proj}_{S}(\mathcal{S})$, $\rank \text{proj}_A (\mathcal{S})$, and $\rank \text{proj}_{AS}(\mathcal{S})$ can be computed without explicitly tracking each generator's Pauli content on $E$. To compute $H_{AS}$ we simply use Eq.~(\ref{eq:EE_stab_1}) with $R = AS$.  Finally, to compute $H_S$ we regard the stabilizer subgroup $\mathcal{S}_S$ as a subgroup of $\mathcal{S}_{AS}$ and apply Eq.~(\ref{eq:EE_stab_2}) to obtain $H_S = N - \abs{\mathcal{G}_{AS}} + \rank \text{proj}_A(\mathcal{S}_{AS})$.

In the case of a globally pure initial state (case $(iii)$), we cannot keep track of the entire stabilizer group $\mathcal{G}$, since $d = N + k + N_E$ increases with each additional system-environment swap. Instead, there is a significantly simpler strategy: since $\rho_t$ is pure at all times, we can compute $H_E = H_{AS}$, $H_{AE} = H_S$, and $H_{SE} = H_A$ using the reduced density matrix $\rho^{AS}_t$ alone. We therefore trace out each swapped-out qubit by keeping track of only the generators $\mathcal{G}_{AS}$.

\begin{figure}
    \centering
    \includegraphics[width = 0.6\textwidth]{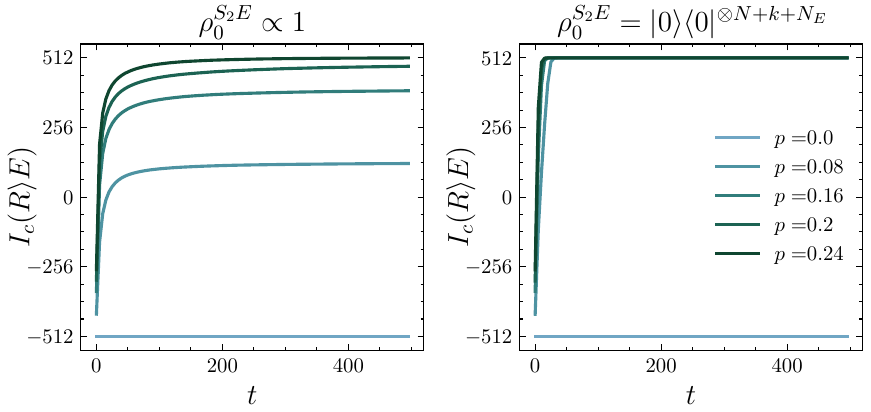}
    \caption{Coherent information $I_c(A \rangle E) = H_E - H_{AE}$ for $S_2 E$ initially maximally mixed(left, case $(i)$) and for $S_2E$ initially pure (right, case $(iii)$), for several swap rates, as a function of time, with $N = k = 512$.}
    \label{fig:info_dynamics}
\end{figure}

Fig.~\ref{fig:info_dynamics} depicts the time evolution of $I_c(A \rangle E)$ for cases $(i)$ and $(iii)$. In the case of an initial maximally mixed environment, the coherent information exhibits the same transition as the OTOC and decoding fidelity in the main text: for $p < p_c$ the coherent information saturates at a value smaller than its maximum value of 512, while for each $p > p_c$ it saturates at the maximum value. In contrast, the coherent information in an initially pure environment rapidly saturates at its maximum value for any nonzero swap rate.

\section{Details of Decoding Protocol}
In this section we derive the relation between the decoder fidelity and the survival probability of the associated DP process discussed in the main text. The expression for the fidelity $\mathcal{F}(t)$ [Eq.~(3) of the main text] is second-order in $U \otimes U^*$; as a result, the Haar and Clifford unitary circuits will yield identical results on average for $\overline{\mathcal{F}(t)}$. We will leverage the technical simplicity of Clifford dynamics in the following derivation, although the same result can be obtained directly using the Haar ensemble.

We consider an initial state $\rho_0$ on $ASS'E$ in which $k$ Bell pairs are shared between $A$ and the first $k$ qubits of $S$, denoted $S_1$. The remaining $N-k$ qubits in $S$ (denoted $S_2$), all $N_E$ environment qubits, and the $N$ ``decoding qubits'' $S' = S_1' \cup S_2'$, are initialized in any stabilizer state of any purity\footnote{Although we assume that $S_2S'E$ is initialized in a stabilizer state, note that the resulting decoding fidelity is linear in $\rho^{S_2 S' E}_0$. As such, the initial state of $S_2 S' E$ can be any state which can be written as a linear combination of stabilizer states with arbitrary single-site rotations. This includes the case of an arbitrary initial product state, as assumed in the main text.}. Explicitly, $\rho_0$ is altogether a stabilizer state of the form
\begin{equation}
\label{eq:rho_0}
	\rho_0 = \frac{1}{2^{M}} \sum_{g \in \mathcal{S}} g = \frac{2^{2k+d}}{2^{M}} \prod_{i = 1}^k \qty[ \qty( \frac{1 + X_{A,i} X_i}{2} ) \qty( \frac{1 + Z_{A,i} Z_i}{2} ) ] \prod_{a = 1}^d \qty(\frac{1 + g_a}{2}) .
\end{equation}
Here $M = 2N + k + N_E$ is the total number of qubits on $ASS'E$, $X_{A,i}$ and $Z_{A,i}$ denote Pauli operators on the $i$th reference qubit, and $\mathcal{S}$ is a stabilizer group generated by the $2k$ stabilizers $X_{A,i} X_i$ and $Z_{A,i} Z_i$ encoding $k$ Bell pairs between $A$ and $S_1$, along with any $d$ additional generators $g_a$. The first $2k$ generators of $\mathcal{S}$ generate a subgroup $\mathcal{S}_{\text{Bell}}$ of $\mathcal{S}$, and the remaining $d$ generators have no support on $A$ or $S_1$. After evolving $SE$ via the unitary circuit $U_t$ and scrambling quantum information initially localized on $S_1$ throughout $SE$, we attempt to decode from the contents of $E$ alone by evolving $S' E$ backwards by $U_t^{\dagger}$; see Fig.~2(a) of the main text.

Upon evolving by the unitary circuit $U_t$, each of the stabilizers with support on $SE$ will both grow within $S$ and swap between $S$ and $E$. At time $t$ we trace out all qubits in $S$, eliminating all stabilizers $g(t) = U_t g U_t^{\dagger}$ with support on $S$:
\begin{equation}
	\rho^{AS'E}_t \equiv \tr_S \rho_t = \frac{2^N}{2^{M}} \sum_{g \in \mathcal{S} : g(t) \eval_{S} = \mathds{1}} g(t) .
\end{equation}
We next evolve backwards by $U_t'^{\dagger}$ to obtain a density matrix $\rho^{AS'E}_{2t}$ given by a sum over stabilizers $g(2t) = U_t'^{\dagger} U_t g U_t^{\dagger} U_t'$. Crucially, any stabilizer $g(2t)$ originating from $\mathcal{S}_{\text{Bell}}$ which survives the trace on $S$ at time $t$ will evolve back to its original form $g(0)$, except with its support on $S$ replaced by $S'$. Upon tracing over $S_2' E$ and computing the expectation value of $\rho^{AS'_1}_{2t} = \tr_{S_2' E} \rho^{AS'E}_{2t}$ with respect to $\ket{\Phi^+_{A S_1'}}$, all stabilizers $g(2t)$ with $g \in \mathcal{S}_{\text{Bell}}$ have eigenvalue $+1$ while any other stabilizers vanish. We therefore obtain the fidelity
\begin{equation}
\label{eq:F}
	\mathcal{F}(t) = \bra{\Phi^+_{AS'_1}} \rho^{AS_1'}_{2t} \ket{\Phi^+_{AS'_1}} = \frac{1}{2^{2k}} \sum_{g \in \mathcal{S}_{\text{Bell}} : g(t) \eval_S = \mathds{1}} 1 .
\end{equation}
Upon averaging over Clifford unitary realizations, we obtain for each $g \in \mathcal{S}_{\text{Bell}}$ the probability for $g$ to evolve to a stabilizer with no support on $S$. This is simply $1-P_g(t)$, where $P_g(t)$ is the survival probability for the associated directed percolation process with an initial distribution of particles specified by the Pauli content of the stabilizer $g$. Altogether,
\begin{equation}
	\overline{\mathcal{F}(t)} = 1 - \frac{1}{2^{2k}} \sum_{g \in \mathcal{S}_{\text{Bell}}} P_g(t) .
\end{equation}
In the particular case $k = 1$, $\mathcal{S}_{\text{Bell}}$ consists of the four stabilizers $\mathds{1}$, $X_{A,1} X_1$, $Z_{A,1} Z_1$, and $-Y_{A,1} Y_1$. The first of these trivially has survival probability zero, and the remaining three have identical survival probability $P_1(t)$ corresponding to an initial condition with a single particle. This gives the result $\overline{\mathcal{F}(t)} = 1 - \frac{3}{4} P_1(t)$ quoted in the main text. For larger $k \geq 1$ we expect each survival probability to undergo the same phase transition at the same swap rate. To obtain bounds on $\overline{\mathcal{F}(t)}$, we note that each survival probability $P_g(t)$ for $g \neq \mathds{1}$ is bounded below by the survival probability $P_1(t)$ for a single initial particle, and is bounded above by the survival probability $P_k(t)$ for an initial condition with $k$ initial particles arranged side-by-side. Since there are $2^{2k} - 1$ nontrivial stabilizers in $\mathcal{S}_{\text{Bell}}$, we obtain the bounds
\begin{equation}
	1 - \frac{2^{2k} - 1}{2^{2k}} P_k(t) \leq \overline{\mathcal{F}(t)} \leq 1 - \frac{2^{2k} - 1}{2^{2k}} P_1(t) ,
\end{equation}
as given in the main text.

\begin{figure}[h]
    \centering
    \hspace{-4.5cm}
    \begin{subfigure}[b]{0.45\textwidth}
        \includegraphics[scale = 0.8]{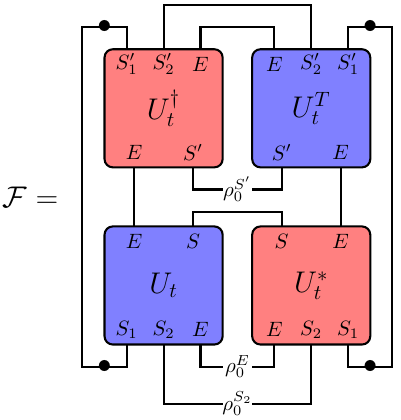}
    \end{subfigure}
    \begin{subfigure}[b]{0.45\textwidth}
        \includegraphics[scale = 0.8]{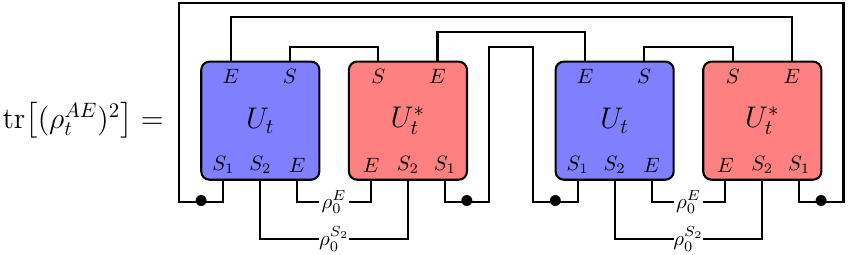}
        \vspace{0.5cm}
    \end{subfigure}
    \caption{Left: tensor network diagram for the decoding fidelity $\mathcal{F}$, for a given circuit $U_t$ and given initial states $\rho^{S_2}_0$, $\rho^E_0$, and $\rho^{S'}_0$. Dots denote factors of $2^{-k/2}$, arising from the Bell states $\ket{\Phi^+_{AS_1}}$. Right: tensor network diagram for the purity $\tr [(\rho^{AE}_t)^2]$, for a given circuit $U_t$ and initial states $\rho^{S_2}_0$ and $\rho^E_0$. When $\rho^{S'}_0$, $\rho^{S_2}_0$, and $\rho^{E}_0$ are each maximally mixed (ie, proportional to $\mathds{1}$,) the two tensor network diagrams can be deformed into each other up to constant factors.}
    \label{fig:fidelity_purity}
\end{figure}

It is useful to note that when $S_2$ and $E$ are initialized in the maximally mixed state, there is a simple relation between the fidelity and the purity of subsystem $AE$ prior to decoding, $\tr [(\rho^{AE}_t)^2]$. This is simply the case in which there are no additional stabilizers $g_a$ (ie $d = 0$), so that $\mathcal{S} = \mathcal{S}_{\text{Bell}}$. As before, tracing $\rho_t$ over $S$ eliminates all stabilizers with nontrivial support in $S$. The remaining stabilizers form a subgroup $\mathcal{S}_{AE} = \qty{ g \in \mathcal{S} : g(t) \eval_{S} = \mathds{1} }$ of $\mathcal{S}$. Squaring the resulting density matrix $\rho^{AE}_t$, we obtain,
\begin{equation}
\label{eq:rho_squared}
	\qty(\rho_t^{AE})^2 = \qty[ \frac{1}{2^{k+N_E}} \sum_{g(t) \in \mathcal{S}_{AE}} g(t)  ]^2 = \frac{1}{2^{2(k+N_E)}} \sum_{g(t) \in \mathcal{S}_{AE}} g(t) \sum_{h(t) \in \mathcal{S}_{AE}} h(t) = \qty[\frac{1}{2^{k+N_E}} \sum_{g (t) \in \mathcal{S}_{AE}} 1] \rho^{AE}_t ,
\end{equation}
where we have reindexed the second sum $g(t) h(t) \to h(t)$. Comparing with (\ref{eq:F}), we find:
\begin{equation}
\label{eq:purity_fidelity}
	\tr [(\rho^{AE}_t)^2] = 2^{k - N_E} \mathcal{F} .
\end{equation}
Using the flatness of the stabilizer spectrum \cite{nahum_quantum_2017,li_measurement-driven_2019,hamma_bipartite_2005,hamma_ground_2005} along with $H_E = N_E$, we obtain the coherent information for a given Clifford circuit realization:
\begin{equation}
\label{eq:info_fidelity}
	I_c(A \rangle E) \equiv H_E - H_{AE} = k + \log_2 \mathcal{F} .
\end{equation}
The transition in the average $\overline{\mathcal{F}}$ suggests that $I_c(A \rangle E)$ exhibits a qualitatively similar transition. We caution, however, that $\log_2 \overline{\mathcal{F}} \neq \overline{\log_2 \mathcal{F}}$, and so $\overline{I_c(A \rangle E)}$ does not have a straightforward statistical mechanics interpretation in either the Haar or Clifford circuits.

Finally, if we further choose $S'$ to be initialized in the maximally mixed state, then Eq.~(\ref{eq:purity_fidelity}) holds more generally for \textit{any} unitary circuit $U_t$ with the given initial conditions. This can be understood visually by comparing the tensor network diagrams for the two expressions, shown in Fig. \ref{fig:fidelity_purity}. The relation (\ref{eq:info_fidelity}) then holds in a particular circuit realization upon replacing $I_c(A \rangle E)$ with its second R\'enyi counterpart, $I^{(2)}_c(A \rangle E) = H^{(2)}_E - H^{(2)}_{AE}$ with $H^{(2)}_R = -\log \tr[(\rho^R_t)^2]$ the second R\'enyi entropy of region $R$. Furthermore, noting that $H^{(2)}_E = H_E = N_E$, as well as $H^{(2)}_{AE} \leq H_{AE}$ and Jensen's inequality \cite{jensen_sur_1906}, we obtain a bound on the average coherent information in terms of the average fidelity:
\begin{equation}
	\overline{I_c(A \rangle E)} \leq H_E^{(2)} - H_{AE}^{(2)} = k + \overline{\log_2 \mathcal{F}} \leq k + \log_2 \overline{\mathcal{F}} .
\end{equation}
This bound is useful in the percolating phase when $\overline{\mathcal{F}} < 1$, so that the average coherent information into the environment is upper-bounded.

\section{Pure-State Decoder with Postselection}
In this section, we describe an alternative decoding protocol which can be used to recover the initial state of $S_1$ on $S_1'$ with perfect fidelity for all $p > 0$, in the special case for which $\rho_0$ is globally pure. The downside of this decoder is that it will require an extensive amount of postselection to achieve the desired outcome: in particular, we will show that this decoding protocol succeeds with probability exponentially small in both $N$ and $k$. 

First, let us provide some intuition for the strong sensitivity of the coherent information $I_c(A \rangle E)$ on the initial state of the swapped-in qubits $E$. As mentioned in the main text, when the initial state of $E$ is maximally mixed, one should regard Eve as having no knowledge of $E$'s initial state; in contrast, if the initial state of $E$ is a pure state, one can regard Eve as having perfect knowledge of this initial state. In the latter case, Eve can use her additional information to design a decoder which succeeds with high fidelity for all nonzero swap rates. 

\begin{figure}
	\centering
	\includegraphics[scale = 0.8]{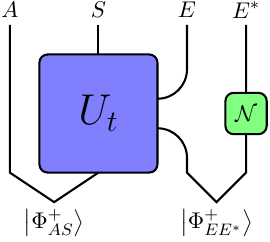}
	\caption{Circuit diagram for a reformulation of the information of the information decoding problem of the main text, in which a classical register $E^*$ keeps track of the initial state of $E$. Tracing over $E^*$ returns the case of a maximally mixed environment, while allowing Eve access to $E^*$ returns the case of an initially pure environment.}
	\label{fig:auxE}
\end{figure}

To see this concretely, let us consider the following modification to the decoding problem: rather than initializing $E$ in a fixed pure or mixed state, we introduce another auxiliary system $E^*$ which is initially maximally entangled with $E$, as in Fig.~\ref{fig:auxE}.  At the end of the evolution we apply a dephasing channel $\mathcal{N}$ to $E^*$, converting it to a classical register which records the initial state of $E$:
\begin{equation}
	\rho_t = \frac{1}{2^{N_E}} \sum_{\vec{x}} \rho^{ASE}_t(\vec{x}) \otimes \dyad{\vec{x}}_{E^*} ,
\end{equation}
where $\vec{x}$ is a bit string of length $N_E$, and $\rho^{ASE}_t(\vec{x}) = U_t \qty[ \dyad{\Phi^+_{AS}} \otimes \dyad{\vec{x}}_E ] U_t^{\dag}$ is the reduced density matrix on $ASE$ in which $E$ is initialized in the state $\ket{\vec{x}}$. 

Tracing over $E^*$ returns the case in which $E$ is initially maximally mixed, so that $I_c(A \rangle E)$ exhibits a transition at $p_c$; on the other hand, we can immediately show that $I_c(A \rangle E E^*)$ is simply the coherent information when $E$ is initialized in a fixed pure state:
\begin{equation}
	I_c(A \rangle E E^*) = H_{E E^*} - H_{A E E^*} = H_{E | E^*} - H_{AE | E^*} = \frac{1}{2^{N_E}} \sum_{\vec{x}} I_c(A \rangle E | \vec{x}) = I_c(A \rangle E | \vec{0}) ,
\end{equation}
where $H_{R | R'} \equiv H_{RR'} - H_{R'}$ is the conditional von Neumann entropy, and $I_c(A \rangle E | \vec{x})$ is the coherent information from $A$ to $E$ in the state $\rho^{ASE}_t(\vec{x})$; note that in our random circuit model, $I_c(A \rangle E | \vec{x})$ is independent of $\vec{x}$, giving $I_c(A \rangle E E^*) = I_c(A \rangle E | \vec{0})$. 

We therefore find that the case in which $E$ is initially a pure state is equivalent to Eve having access to the classical register $E^*$. In this language it is clear that Eve's ability to recover Alice's information is dramatically stronger when $E$ is initially pure, since Eve has access to $N_E \sim p N t$ additional bits of information. 

We now demonstrate our alternative protocol for an initially pure environment. We assume as before that $A$ and $S_1$ initially share $k$ Bell pairs, but in addition we suppose $\rho_0^{S_2 E} = \dyad{0}^{\otimes N-k+N_E}$ is initialized in a pure product state. The decoding protocol proceeds initially as before: after evolving $SE$ by the unitary circuit $U_t$, Eve attempts to decode Alice's state from the contents of $E$ by introducing an extra set of $N$ qubits $S'$, and evolving $S' E$ backwards via $U_t^{\dagger}$. At this stage, rather than tracing out $S_2' E$, Eve instead performs projective measurements on $S_2' E$ and postselects on the outcome $\dyad{0}^{\otimes N-k+N_E}$. We will argue that these measurements have the effect of teleporting Alice's state onto $S_1'$.

\begin{figure}[h]
    \centering
    \hspace{-4.5cm}
    \begin{subfigure}[b]{0.45\textwidth}
        \includegraphics[scale = 0.7]{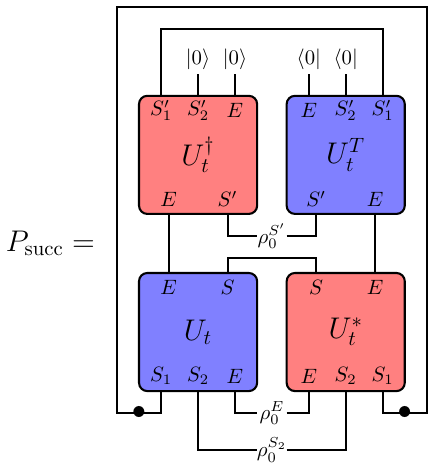}
    \end{subfigure}
    \begin{subfigure}[b]{0.45\textwidth}
        \includegraphics[scale = 0.7]{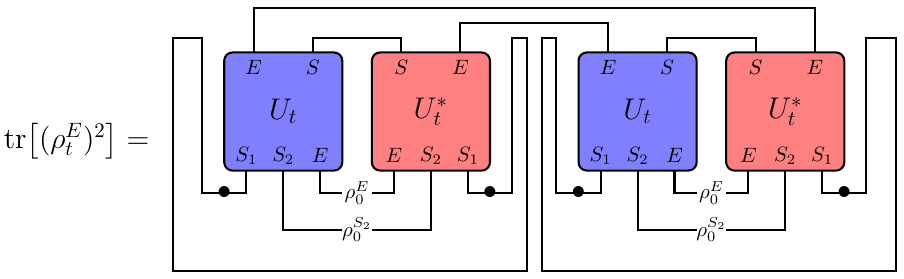}
        \vspace{0.5cm}
    \end{subfigure}

    \hspace{-2.5cm}
    \begin{subfigure}[b]{0.45\textwidth}
        \includegraphics[scale = 0.7]{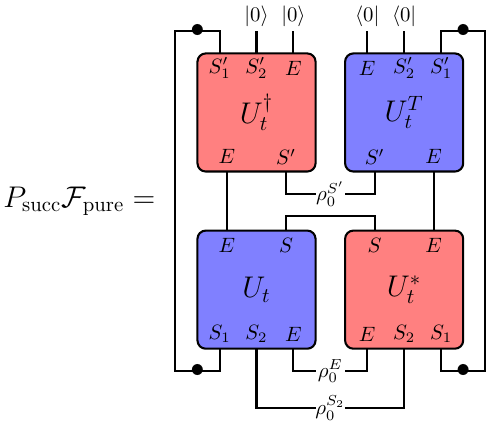}
    \end{subfigure}
    \caption{Left: tensor network diagram for the success probability $P_{\text{succ}}$ of achieving the desired postselection outcome. Dots indicate factors of $2^{-k/2}$, arising from the Bell state $\ket{\Phi^+_{AS_1}}$. Right: tensor network diagram for the purity $\tr [(\rho^E_t)^2]^2$, for a given circuit $U_t$ and initial states $\rho_0^{S_2}$ and $\rho^E_0$. When $\rho^{S_2}_0 = \dyad{0}^{\otimes N-k}$ and $\rho^E_0 = \dyad{0}^{\otimes N_E}$, the middle diagram can be deformed into the left diagram up to constant numerical factors. Bottom: tensor network diagram for the decoding fidelity $P_{\text{succ}} \mathcal{F}_{\text{pure}}$ in the pure-state postselection decoding scheme. When $\rho^{S_2}_0 = \dyad{0}^{\otimes N-k}$ and $\rho^E_0 = \dyad{0}^{\otimes N_E}$, this diagram and the right diagram of Fig.~\ref{fig:fidelity_purity} can be deformed into each other.}
    \label{fig:pure_decoder}
\end{figure}

We first determine the probability $P_{\text{succ}}$ of successful measurements on $S_2' E$, given by the tensor network in the leftmost figure of Fig.~\ref{fig:pure_decoder}. By deforming the tensor network it can be related to the purity $\tr [(\rho^E_t)^2]$ of subsystem $E$ prior to decoding, given by the righthand figure of Fig.~\ref{fig:pure_decoder}. In particular, we have
\begin{equation}
    P_{\text{succ}} = \bra{0_{S_2' E}} \rho_{2t}^{S_2' E} \ket{0_{S_2' E}} = \frac{2^k}{2^N} \tr [(\rho^E_t)^2] ,
\end{equation}
where $\ket{0_{S_2' E}} = \ket{0}^{\otimes N-k+N_E}$. We can roughly estimate the purity of $E$ by noting that it is identical to the purity of $AS$, and by assuming $\rho^{AS}_t \simeq \rho^A_t \otimes \rho^S_t$ rapidly factorizes for typical circuits as in the main text. The reference $A$ has minimal purity $\tr[(\rho^A_t)^2] = 2^{-k}$ at all times, while we estimate $\tr \qty[ (\rho^S_t)^2 ] \simeq 2^{-(1-p)N}$ by assuming $\rho^S_t$ has $pN$ pure swapped-in qubits at each time-step while the remaining qubits are maximally entangled with $E$ at sufficiently late times. We therefore estimate that $P_{\text{succ}}$ is exponentially small in $(1-p)N + k$.

We now show that the fidelity of this decoding protocol, given successful postselection, is related to the coherent information $I_c(A \rangle E)$ into the environment qubits. The fidelity within this protocol is given by
\begin{equation}
    \mathcal{F}_{\text{pure}} = \frac{1}{P_{\text{succ}}} \bra{\Phi^+_{AS_1'}, 0_{S_2' E}} \rho^{AS'E}_{2t} \ket{\Phi^+_{AS_1'}, 0_{S_2' E}} ,
\end{equation}
where the factor of $P_{\text{succ}}$ in the denominator arises from the normalization of the density matrix following the measurement. The tensor network for $P_{\text{succ}} \mathcal{F}_{\text{pure}}$ is given in the righthand figure of Fig.~\ref{fig:pure_decoder}. As in the original decoder, we can deform this tensor network to match that of the purity $\tr [(\rho^{AE}_t)^2]$ prior to decoding, yielding the result
\begin{equation}
    \mathcal{F}_{\text{pure}} = \frac{1}{2^k} \frac{\tr[(\rho^{AE}_t)^2]}{\tr [(\rho^E_t)^2]} = 2^{I_c(A \rangle E) - k} .
\end{equation}
If we once again assume $\rho^{AS}_t \simeq \rho^A_t \otimes \rho^S_t$ rapidly factorizes in typical circuits, then we obtain $I_c(A \rangle E) = -I_c(A \rangle S) = k$, giving the maximal fidelity $\mathcal{F} = 1$.

\section{Absence of Percolating Phase in a Non-Scrambling Model}
In this section, we demonstrate the importance of scrambling unitary dynamics in obtaining a phase transition at swap rates $p > 0$. Specifically, we show that if the Haar or Clifford random unitary gates are replaced by free fermion unitary evolution, then all local operators will leak into the environment exponentially quickly for any nonzero swap rate.

To be specific, consider a system of $N$ Majorana fermions $\gamma_j$ satisfying the anticommutation relations $\acomm{\gamma_i}{\gamma_j} = 2\delta_{ij}$, evolving under a free fermion Hamiltonian $H$ given by
\begin{equation}
\label{eq:free_fermion_ham}
	H = \frac{i}{4} \sum_{ij}^{N} A_{ij} \gamma_i \gamma_j .
\end{equation}
The real antisymmetric matrix $A_{ij}$ may exhibit local or nonlocal inter-system interactions; it can also be chosen to be time-dependent or to have random elements, but for simplicity we take $A_{ij}$ to be fixed and time-independent. At time steps $t = n \delta t$ ($n \in \mathbb{Z}$), we interrupt the unitary dynamics and interact each $j$th fermion with an auxillary fermion $\eta_{j,n}$ via the unitary gate
\begin{equation}
	u_{j,n} = e^{ \theta_{j,n} \gamma_j \eta_{j,n}} .
\end{equation}
We can choose each $\theta_{j,n} \in [0,2\pi)$ to be a fixed number, or randomly distributed; for sake of concreteness, we will take $\theta_{j,n} \equiv \theta$ to be fixed in time and space, but this feature is not essential for the physics to follow. 

The crucial feature of free fermion dynamics, as compared to generic dynamics, is that the Heisenberg evolution of a single-site fermion operator remains a linear combination of single-site fermion operators at all times. Specifically, the Heisenberg evolution of the operator $\gamma_i$ by a time step $\delta t$ via the Hamiltonian (\ref{eq:free_fermion_ham}) yields
\begin{equation}
	U_{\delta t}^{\dag} \gamma_i U_{\delta t} = \sum_{j = 1}^N [e^{A \delta t}]_{ij} \gamma_j .
\end{equation}
As a result, an initially local fermion operator $\gamma_0$ evolved to time $t$ can be expanded as a linear combination of single-site operators:
\begin{equation}
	\gamma_0(t) \equiv U_t^{\dag} \gamma_0 U_t = \sum_{j = 1}^N w_j(t) \gamma_j + \sum_{j, n}^{N_E} \bar{w}_{j,n}(t) \eta_{j,n} .
\end{equation}
This expansion should be contrasted with the analogous expansion for operator strings in Sec.~\ref{sec:OTOC_qudit}; whereas $X^a_0(t)$ in a scrambling unitary circuit is generically a linear combination of $2^M$ possible operator strings $X^{\vec{c}}$ (many of which are highly nonlocal), the operator $\gamma_0(t)$ under free fermion evolution is restricted to a linear combination of $M$ single-site operators. This is closely related to the fact that free fermion dynamics are restricted to exploring a vastly smaller region of the total Hilbert space than generic many-body dynamics.

In the case of isolated dynamics (i.e., $\theta = 0$), the total weights $w_j(t)$ within the system are conserved: $\sum_j [w_j(t)]^2 = 1$. On the other hand, for $\theta \neq 0$, $\gamma_0(t)$ will develop support on the environment fermions $\eta_{j,n}$. Specifically, 
\begin{equation}
	u_{j,n}^{\dag} \gamma_j u_{j,n} = \gamma_j \cos(2\theta) + \eta_{j,n} \sin(2\theta) .
\end{equation}
As a result, any weight $w_j(t)$ prior to the unitary $u_{j,n}$ evolves to a weight $w_j(t) \cos(2\theta)$ following $u_{j,n}$. The total weight on all system fermions then simply evolves as
\begin{equation}
	\sum_{j = 1}^N [w_j(t)]^2 = [\cos^2(2\theta)]^{t/\delta t} .
\end{equation}
We immediately find, under a very broad class of free fermion dynamics, that the operator weight of $\gamma_0(t)$ within the system decays exponentially quickly in time. As a result, there is no ``percolating phase'' in a radiative free fermion model; all local operators (and thereby all initially localized quantum information) are lost to the environment within a finite timescale.

\bibliographystyle{apsrev4-2}
\bibliography{supp_refs}